\documentclass[useAMS,usenatbib]{mn2e}
\usepackage{times}
\usepackage{graphicx}
\usepackage{rotating}

\title[Balloon\,090100001: 2005 campaign]{The pulsating hot subdwarf Balloon 090100001: results of the 2005 multisite campaign}
\author[A.\,Baran et al.]
{A.\,Baran$^{1,2}$\thanks{E-mail:andy@astro.as.ap.krakow.pl}, R.\,Oreiro$^{3,4}$, A.\,Pigulski$^{5}$, 
F.\,P{\'e}rez Hern\'andez$^{3,6}$, A.\,Ulla$^{7}$, M.D.\,Reed$^{8}$,\newauthor
C.\,Rodr\'iguez-L\'opez$^{7,9,10}$, P.\,Moskalik$^{11}$, S.-L.\,Kim$^{12}$, W-P.\,Chen$^{13}$, R.\,Crowe$^{14}$,\newauthor
M.\,Siwak$^{15}$, L.\,Armendarez$^{14}$, P.M.\,Binder$^{14}$, K.-J.\,Choo$^{12}$, A.\,Dye$^{14}$, J.R.\,Eggen$^{8}$,\newauthor
R.\,Garrido$^{9}$, J.M.\,Gonz\'alez P\'erez$^{3,16}$, S.L.\,Harms$^{8}$, F.-Y.\,Huang$^{13}$, D.\,Kozie{\l}$^{15}$,\newauthor
H.-T.\,Lee$^{13}$, J.\,MacDonald$^{17}$, L.\,Fox~Machado$^{3,18}$, T.\,Monserrat$^{3}$, J.\,Stevick$^{14}$,\newauthor
S.\,Stewart$^{14}$, D.\,Terry$^{14}$, A.-Y.\,Zhou$^{8,19}$, S.\,Zo{\l}a$^{1,15}$ \\
$^{1}$Cracow Pedagogical University, ul.\,Podchor\c{a}\.zych\,2, 30\,--\,084 Krak\'ow, Poland \\
$^{2}$Toru\'n Centre for Astronomy, ul.\,Gagarina\,11, Toru\'n, Poland \\
$^{3}$Instituto de Astrof\'isica de Canarias, 38200 La Laguna, Spain \\
$^{4}$Institue of Astronomy, Katholieke Universiteit Leuven, Celestijnenlaan\,200D, 3001 Leuven, Belgium \\
$^{5}$Instytut Astronomiczny Uniwersytetu Wroc{\l}awskiego, ul.\,Kopernika\,11, 51\,--\,622 Wroc{\l}aw, Poland \\
$^{6}$Departamento de Astrof\'isica, Universidad de La Laguna, 38200 La Laguna, Spain \\
$^{7}$Departamento de F\'isica Aplicada, Universidade de Vigo, 36200 Vigo, Spain \\
$^{8}$Department of Physics, Astronomy and Materials Science, Missouri State University, 901\,S.\,National, Springfield, MO 65897, USA \\
$^{9}$Instituto de Astrof\'isica de Andaluc\'ia\,--\,C.S.I.C., Camino Bajo de Hu\'etor\,50, 18008 Granada, Spain \\
$^{10}$Laboratoire d'Astrophysique de Toulouse-Tarbes, Universit\'e de Toulouse, CNRS, 14 Av.~Edouard Belin, Toulouse 31400, France\\
$^{11}$Copernicus Astronomical Centre, ul.\,Bartycka 18, 00-716 Warsaw, Poland\\
$^{12}$Korea Astronomy and Space Science Institute, Daejeon 305-348, South Korea \\
$^{13}$Institute of Astronomy, National Central University, Jhongli 32054, Taiwan \\
$^{14}$Department of Physics and Astronomy, University of Hawaii\,--\,Hilo, 200\,West Kawili Street, Hilo, Hawaii 96720-4091, USA \\
$^{15}$Astronomical Observatory, Jagiellonian University, ul.\,Orla\,171, 30-244\,Krak\'ow, Poland\\
$^{16}$GRANTECAN S.A.~(CALP), 38712 Brena Baja, La Palma, Spain\\
$^{17}$Departament of Physics and Astronomy, University of Delaware, Newark, DE 19716, USA\\
$^{18}$Observatorio Astron\'omico Nacional, Instituto de Astronom\'{\i}a, Universidad Aut\'onoma de M\'exico, A.P.\,877 Ensenada, BC 22860, Mexico\\
$^{19}$National Astronomical Observatories of the Chinese Academy of Sciences, Beijing 100012, P.R.~China}
\begin{document}
\date{Accepted .... Received ....; in original form ....}
\pagerange{\pageref{firstpage}--\pageref{lastpage}} 
\pubyear{2008}
\label{firstpage}
\maketitle

~

~

~

~

~

~

~

~

~

~

~

~

~

~

~

~

~

~

~

~

~

~

~

~

~

~

~

~

~

~

~

~

~

~

~

~

~

~

~

~

~

~

~

~

~

~

~

~

\begin{abstract}
We present the results of a multisite photometric campaign on the pulsating sdB star Balloon\,090100001. 
The star is one of the two known hybrid hot subdwarfs with both long- and short-period oscillations,
theoretically attributed to $g$- and $p$-modes. The campaign involved eight telescopes with
three obtaining $UBVR$ data, four $B$-band data, and one Str\"omgren $uvby$ photometry.
The campaign covered 48 nights, providing a temporal resolution of 0.36\,$\mu$Hz with a detection
threshold of about 0.2\,mmag in $B$-filter data.

Balloon\,090100001 has the richest pulsation spectrum of any known pulsating subdwarf B star and our analysis
detected 114 frequencies including 97 independent and 17 combination
ones. Most of the 24 $g$-mode frequencies are between 0.1 and 0.4\,mHz.
Of the remaining 73, presumably $p$-modes, 72 group into four distinct regions near 2.8, 3.8, 4.7 and 5.5\,mHz.
The density of frequencies requires that some modes must have degrees $\ell$ larger than 2.
The modes in the 2.8~mHz region have the largest amplitudes. The strongest mode ($f_1$) is most likely radial while
the remaining ones in this region form two nearly symmetric multiplets: a triplet and quintuplet, attributed
to rotationally split $\ell$ = 1 and 2 modes, respectively. We find clear increases of splitting in
both multiplets between the 2004 and 2005 observing campaigns, amounting to $\sim$15\% on average. 
The observed splittings imply that the rotational rate in Bal09 depends on stellar latitude and is the fastest 
on the equator. We also speculate on the possible reasons for the changes of splitting. The only plausible 
explanation we find are torsional oscillations.  This hypothesis, however, need to be verified in the future 
by detailed modelling. In this context, it is very important to monitor the splittings on a longer time scale 
as their behaviour may help to explain this interesting phenomenon.

The amplitudes of almost all terms detected both in 2004 and 2005 were found to vary. This is evident even
during one season; for example, amplitudes of modes $f_8$ and $f_{\rm C}$ were found to change by a factor of 2--3 
within about 50 days during 2005. 

We use a small grid of models to constrain the main mode ($f_1$), which most likely represents the radial
fundamental pulsation. The groups of $p$-mode frequencies appear to lie in the vicinity of consecutive
radial overtones, up to the third one. Despite the large number of $g$-mode frequencies observed, we failed to 
identify them, most likely because of the disruption of asymptotic behaviour by mode trapping.  
The observed frequencies were not, however, fully exploited in terms of seismic analysis which should be
done in the future with a larger grid of reliable evolutionary models of hot subdwarfs.
\end{abstract}


\vfill\eject
\section{Introduction}
\citet{kilkenny97} discovered the first group of B-type hot subdwarf (sdB) pulsators, which
they termed EC\,14026 after the prototype and are now designated as V\,361\,Hya stars, and referred to as 
sdBV stars.
This discovery opened the possibility of constraining the interiors of sdB stars through
asteroseismology, which, in turn, may help to better understand their still controversial evolution.
So far, about 40 class members have been found, with periods ranging from 60 to 600\,s and typical
amplitudes of 10\,mmag, with some exceptions of higher amplitudes up to 60\,mmag. Theoretical studies by
\cite{charpinet97} demonstrated that their oscillations could be explained by acoustic ($p$-) modes of
low degree, $\ell$, and low radial orders, $n$, driven by an opacity bump caused by a local enhancement of 
iron-group elements.

In 2002 another class of pulsating sdB stars was found \citep{green03}. The new class
is characterised by lower amplitudes (1\,--\,2\,mmag) and longer periods ($\sim$1\,hour). The new variable 
stars are referred to as long-period sdBs, or PG\,1716 stars, after that prototype. \cite{fontaine03}
determined that their variability could be explained in terms of high radial order gravity ($g$-) modes.
The same $\kappa$ mechanism that operates in V\,361~Hya stars also works in PG\,1716 stars,
although theoretical unstable modes were at slightly lower effective temperatures than observed.
\cite{jeffery06}, using OP opacity tables and considering an enhancement of both iron and nickel in the
driving zone, were able to obtain unstable modes at the same effective temperatures as the observed ones.
Additionally, their non-adiabatic analysis led to simultaneous instability of $p$- and $g$-modes, which can account
for the two known hybrid sdB pulsators.

The two classes (V\,361~Hya and PG\,1716 stars) of pulsating sdBs have slightly different effective
temperatures and surface gravities, the former being hotter and denser than the latter. Their pulsating
behaviour resembles two other pairs of pulsating stars showing $p$- and $g$-modes: $\beta$ Cephei and Slowly Pulsating
B stars and $\delta$ Scuti and $\gamma$ Doradus stars.

\section{Balloon\,090100001}
Balloon\,090100001 (hereafter Bal\,09) is one of the brightest pulsating sdB stars discovered to date.
It was found to be an sdB star by \cite{bixler91}, and a short-period pulsating sdB by \cite{oreiro04}. It
has relatively long periods and large amplitudes of oscillation, which makes this object an excellent
candidate for follow-up photometry. Bal\,09 was observed in a long-time base campaign independently by
two of us, in August and September 2004 using the 60-cm telescope at Mt.~Suhora Observatory (AB) and the
80-cm IAC80 telescope at Tenerife (RO). Four wide-band filters, $UBVR$, were used in the former 
\citep[hereafter Bar05]{baran05}, while only $B$-filter data were obtained in the latter \citep[hereafter 
Ore05]{oreiro05}. The use of a common filter allowed an analysis
of the combined data, from which almost 50 frequencies were detected \citep{baran06}. These campaigns revealed
a rich spectrum of frequencies clustering in groups within the $p$-mode domain, and having
lower amplitudes as the frequency increases. A high-amplitude frequency ($\sim$\,60\,mmag) dominates the
spectrum, with an equally-spaced triplet nearby. Assuming that the triplet splitting is caused by rotation,
a rotational period near 7\,days was derived. Interestingly, variability near 4\,mmag  was also
detected in the low-frequency region, which is typical for PG1716 stars. Thus, Bal\,09 turned out
to be a hybrid object and became very interesting from the seismological point of view. The star was
the second hybrid sdB pulsator discovered, the first being HS\,0702+6043 \citep{schuh06}, which shows a single frequency in the
$g$-mode region. 

\section{2005 photometric campaign}
Despite the frequency resolution achieved with the 2004 observations, the spectrum of Bal\,09 suffered from
strong aliasing, hindering a clear frequency identification at lower amplitudes. Closely-spaced frequencies
and/or amplitude variability made the situation even worse.

\begin{table*}
\centering
\caption[]{Details on the 2005 campaign. UH stands for University of Hawaii at Hilo. 
Other observer's abbreviations correspond to initials of the co-authors. }
\label{logobserv}
\begin{tabular}{ccccrrc}
\hline
Telescope& Filter & HJD start  & HJD end    & Nights& Hours    & Observers \\
\hline
IAC~Teide, 0.8\,m     &  $U$   & 620.37 & 628.75 & 7     &  44.2   & RO, JMGP, TM, LF\\
                      &  $B$   & 593.54 & 628.75 & 18    &  115.2  & \\
                      &  $V$   & 620.37 & 628.75 & 7     &  44.2   & \\
                      &  $R$   & 620.37 & 628.75 & 7     &  44.1   & \\
\hline
Mt.~Suhora, 0.6\,m    &  $U$   & 594.47 & 639.51 & 34    &  226.9  & AB, MS, DK, SZ\\
                      &  $B$   & 594.47 & 639.51 & 36    &  216.6  & \\
                      &  $V$   & 594.47 & 639.51 & 34    &  226.9  & \\
                      &  $R$   & 594.47 & 639.51 & 34    &  226.9  & \\
\hline
Baker, 0.4\,m         &  $U$   & 594.68 & 624.91 & 11    &  79.2   & MR, AYZ, SH, JRE\\
                      &  $B$   & 594.68 & 625.87 & 13    &  92.4   & \\
                      &  $V$   & 594.68 & 594.79 & 1     &  2.7    & \\
                      &  $R$   & 594.68 & 624.96 & 11    &  80.0   & \\
\hline
Lulin, 1\,m           &  $B$   & 609.01 & 613.13 & 5     &  21.9   & WPC, HTL, FYH\\
\hline
Sobaeksan, 0.6\,m     &  $B$   & 618.01 & 630.03 & 5     &  23.3   & SLK, KJC\\
\hline
Mauna~Kea, 0.6\,m     &  $B$   & 621.87 & 631.02 & 7     &  36.7   & AB, RC, UH students\\
\hline
Mt.~Lemmon, 1\,m      &  $B$   & 636.84 & 640.63 & 5     &  29.0   & SLK\\
\hline
Sierra Nevada, 0.9\,m & $uvby$ & 605.41 & 618.62 & 14    &  105.6  & CRL, RG\\
\hline
\end{tabular}
\end{table*}

In order to get through these problems, a multisite campaign was organized in the summer of 2005. It involved
eight telescopes at different longitudes, and spanned almost two months (10th August\,--\,27th September).
All sites but one used CCD cameras with Johnson $B$ filter (or very similar), although some of them performed
multicolour photometry in $UBVR$ filters. Moreover, Str\"omgren $uvby$ photometry was acquired with 
a photoelectric photometer attached to the 90-cm telescope in Sierra Nevada Observatory. The log of
observations, including sites, observers and other relevant information is given in Table \ref{logobserv}.

\subsection{Data reduction}
Raw CCD images were calibrated in a standard way by removing instrumental effects using bias, dark
and flat field images. Magnitudes of stars were extracted using aperture photometry, but by means of different
reduction pipelines: data collected at IAC\,80 (Tenerife) were reduced with the Real Time Photometry (RTP)
software \citep{ostensen2000}; IDL photometry packages were applied to the Lulin (Taiwan) data; photometry
from Sobaeksan (Korea) and Mt.\,Lemmon (Arizona) were proccessed with IRAF routines, while other CCD data
were reduced by means of the DAOPHOT package with a DAOGROW supplement \citep{stetson87,stetson90}.
Data from the four-channel photoelectric photometer attached to the 90-cm telescope in Sierra
Nevada Observatory were corrected with nightly extinction coefficients derived from 
a nearby star.

\begin{figure}
\includegraphics[width=83mm]{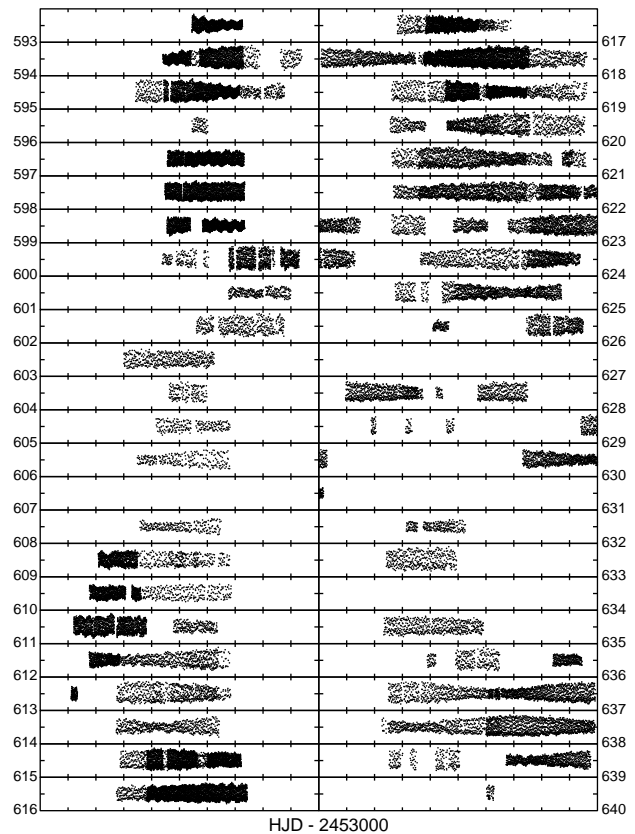}
\caption{\small Observational data of Bal\,09 in $B$ and the average of $b$ and $v$ filters obtained in August
and September 2005. Each panel covers one day and the numbers in the bottom left or right corners represent
the three
last integer digits of HJD for that panel. Note that consecutive days go from top to bottom. The range in ordinate 
amounts to 240~mma.}
\label{data_raw_b_2005}
\end{figure}

Differential photometry for all filters was obtained using the reference star GSC\,02248.00063,
the most suitable due to its location in the field of view and magnitude. However, as it is much cooler than
Bal\,09, all data were corrected for second-order extinction effects. If, after this step, the light
curve still showed long-period variations (on time scales much longer than a few hours), a spline fit was
applied and subtracted from the data to remove those trends. The resulting light curves, obtained
for each site separately, were then shifted to the same mean, converted to relative flux and combined.
A similar conversion to relative fluxes has been done for the 2004 data. Consequently, all amplitudes
provided in this paper are given in the units of 1/1000 of the relative flux (mma).
In Fig.~\ref{data_raw_b_2005}, the entire $B$ and $bv$ averaged light curves are presented.

\subsection{Re-analysis of the combined 2004 data}
As previously described, two independent photometric data sets of Bal\,09 were obtained in 2004.
As our analysis of the 2005 campaign data revealed amplitude variations 
(see Section \ref{ft}), we decided to combine the 2004 $B$-filter
data and re-analyse them in the same way as the 2005 campaign data. The 2004 data of Bar05
are better distributed in time and cover a longer time interval than those of Ore05, allowing for
a better distinction of close frequencies. On the other hand, the data of Ore05 have better precision
and were carried out with shorter time sampling. In order to balance the accuracy and resolution in the
combined data, we decided to average three consecutive datapoints in the data of Ore05 before
combining them together.

\begin{table*}
\centering
\footnotesize
\caption{A list of modes detected in the combined 2004 $B$ filter data, while allowing for linear
amplitude changes. Amplitudes and phases are given for HJD = 2\,453\,246.
Numbers in brackets indicate the formal r.m.s.~errors.}
\begin{tabular}{clrrlclrrl}
\hline\noalign{\smallskip}
 \# & Frequency & Ampl. & d$A$/d$t$ & Phase & \# & Frequency & Ampl. & $dA/dt$ & Phase \\
    &      [mHz]     &  $A$ [mma]  & [mma/d] & [rad] &  &      [mHz]     &  $A$ [mma]  & [mma/d] & [rad] \\
\noalign{\smallskip}\hline\noalign{\smallskip}
$f_{\rm I}$ & 0.229566(24) & 0.57 & $+$0.013 & 0.12(17) &$f_{26}$& 3.822911(21)  & 0.70 & $-$0.002 & 0.14(13) \\
$f_{\rm D}$ & 0.239973(07) & 1.96 & $+$0.010 & 5.80(05) &&&&&\\
$f_{\rm F}$ & 0.246315(22) & 0.61 & $+$0.003 & 2.64(15) &$f_{11-}$& 4.633620(25)  & 0.57 &    0.000 & 1.26(17) \\
$f_{\rm A}$ & 0.272379(05) & 2.75 & $-$0.028 & 5.66(03) &$f_{21}$ & 4.644467(19)  & 1.14 & $-$0.034 & 1.96(10) \\
$f_{\rm G}$ & 0.298886(19) & 0.67 & $+$0.003 & 6.23(13) &$f_{27}$ & 4.651166(25)  & 0.65 & $-$0.005 & 2.20(15) \\
$f_{\rm C}$ & 0.325671(12) & 1.25 & $+$0.052 & 1.37(09) &$f_{28}$ & 4.655560(28)  & 0.80 & $-$0.022 & 4.35(15) \\
$f_{\rm J}$ & 0.331210(16) & 0.81 & $+$0.007 & 1.42(11) &$f_{29}$ & 4.660107(22)  & 0.76 & $-$0.009 & 2.70(13) \\
$f_{\rm B}$ & 0.365807(05) & 2.66 & $-$0.023 & 1.98(03) &$f_{13}$ & 4.661418(14) & 1.07 & $+$0.005 & 2.04(09) \\
$f_{\rm K}$ & 0.397189(26) & 0.49 & $+$0.011 & 0.48(20) &$f_{30}$ & 4.670320(17)  & 0.92 & $-$0.002 & 4.24(11) \\
$f_{\rm L}$ & 0.631078(29) & 0.44 & $+$0.011 & 0.94(22) &$f_{31}$ & 4.676034(34)  & 0.65 & $-$0.026 & 2.57(21) \\
$f_{\rm M}$ & 0.684352(21) & 0.60 & $+$0.002 & 4.77(14) &$f_{12+}$  & 4.680947(21)  & 0.78 & $-$0.006 & 4.00(12) \\
$f_{\rm N}$ & 0.833085(25) & 0.53 & $-$0.006 & 4.84(17) &&&&&\\
&&&&                                                    &$f_{32}$ & 5.506010(28)  & 0.54 & $-$0.013 & 1.16(18) \\
$f_8 - f_1$ & 0.968616     & 0.43 & $+$0.001 & 5.21(20) &$f_{33}$ & 5.532958(21)  & 0.61 & $+$0.009 & 3.86(15) \\
&&&&                                                    &$f_{34}$ & 5.553408(27)  & 0.48 & $-$0.002 & 0.11(18) \\
$f_1 - f_{\rm B}$& 2.441662& 0.80 & $+$0.001 & 1.11(11) &$f_{35}$ & 5.605820(28)  & 0.47 & $+$0.002 & 4.13(19) \\
$f_1 - f_{\rm C}$& 2.481799& 0.43 & $+$0.016 & 2.13(19) &&&&&\\
$f_1 - f_{\rm A}$& 2.535091& 0.41 & $+$0.009 & 3.57(20) & 2$f_1$ & 5.6149394 & 5.67 & $+$0.003 & 3.874(16) \\
&&&&                                                    & $f_1 + f_2$ & 5.6307087     & 4.33 & 0.000    & 1.793(21) \\
$f_1$       & 2.8074697(02) &52.83 & $+$0.058 & 2.6938(17) & $f_1 + f_3$ & 5.6322777 & 2.61 & $-$0.022 & 0.94(04) \\
$f_2$       & 2.8232390(06) &20.39 & $+$0.032 & 0.602(05) & $f_1 + f_4$ & 5.6338393 & 0.99 & $-$0.003 & 4.10(10) \\
$f_3$       & 2.8248080(11) &12.13 & $-$0.055 & 5.997(07) & 2$f_2$      & 5.6464780     & 0.80 & $+$0.001 & 5.75(12) \\
$f_4$       & 2.8263696(27) & 5.09 & $-$0.034 & 2.784(18) & $f_2 + f_3$ & 5.6480470     & 0.66 & $-$0.002 & 4.87(14) \\
$f_5$       & 2.853958(08)  & 1.73 & $-$0.001 & 5.19(5) & $f_1 + f_5$ & 5.661428      & 0.53 & $+$0.002 & 0.44(17) \\
$f_7$       & 2.855710(10)  & 1.38 & $+$0.009 & 2.27(7) &&&&&\\
$f_6$       & 2.858534(14)  & 1.15 & $-$0.002 & 4.86(9) & $f_1 + f_8$ & 6.583556  & 0.55 & $-$0.013 & 1.90(15) \\
 &&&&                                                   &&&&&\\
$f_{15}$    & 3.763702(25)  & 0.61 & $+$0.009 & 1.51(17) & 3$f_1$ & 8.4224091 & 0.71 & $-$0.010 & 5.37(12) \\
$f_{23}$    & 3.764802(30)  & 0.64 & $-$0.003 & 2.54(16) & 2$f_1 + f_2$& 8.4381784     & 0.49 & $+$0.005 & 2.83(18) \\
$f_8$       & 3.776086(05)  & 4.50 & $-$0.045 & 0.914(24) & 2$f_1 + f_3$& 8.4397475     & 0.40 & $-$0.009 & 2.05(21) \\
$f_9$       & 3.786724(13)  & 1.33 & $-$0.007 & 2.51(08) &&&&&\\
$f_{24}$    & 3.791843(18)  & 1.00 & $-$0.028 & 3.61(11) &\multicolumn{3}{r}{N$_{\rm obs}$}& 12115&\\
$f_{10}$    & 3.795572(13)  & 1.05 & $+$0.003 & 1.60(09) &\multicolumn{3}{r}{Detection threshold [mma]}& 0.42 &\\
$f_{18^\prime}$    & 3.797813(23)  & 0.68 & $+$0.007 & 4.94(16) &\multicolumn{3}{r}{$\sigma_A$, $\sigma_{dA/dt}$ [mma, mma/d]}& 0.09 & 0.007\\
$f_{25}$    & 3.821887(28)  & 0.54 & $+$0.014 & 1.22(20) &\multicolumn{3}{r}{Residual S.Dev. [mma]}& 6.46 &\\
\noalign{\smallskip}\hline
\end{tabular}
\label{a_fi_2004}
\end{table*}

Section \ref{ft} explains details of the analysis of the 2005 data. For the sake of consistency, the combined 2004 data 
were analysed in the same way.  The new feature was the inclusion of the
rates of amplitude changes in the non-linear
least squares used to subtract periodic terms found by means of Fourier periodogram.
The rates of these changes as well as the amplitudes and phases of all frequencies detected
in the 2004 data are reported in Table \ref{a_fi_2004}. In total, we detected 56 frequencies of which 41 (12 in $g$-mode and 29 in $p$-mode regions)
are independent and 15 are combination terms. In comparison to the analysis of Bar05,
we found 26 more frequencies. This is mainly the consequence of lowering the detection level in the
combined data. Of the 30 modes given by Bar05, we did not detect $f_{\rm E}$ and $f_{\rm H}$ in the
$g$-mode region. This creates uncertainty as to whether they are intrinsic to Bal\,09, but it is also possible that 
their amplitudes were changing quite fast, making them undetectable in the combined data
\citep[see for example][]{reed07a}. This is only possible if their
amplitudes were very low at the beginning of the 2004 run when most of the Ore05 data were acquired.
It is worth noting that the combined data revealed a fourth region of $p$-mode frequencies
around 5.5~mHz. Frequencies in this region were reported by Ore05, but
not detected by Bar05.

In Table \ref{a_fi_2004} and throughout this paper, the notation of frequencies introduced by 
Bar05 is used and extended for new modes: 
frequencies in the $g$-mode region are annotated by letters while numbers are assigned to frequencies in the
$p$-mode region. When a frequency was recovered at an alias frequency
it is marked with `$+$' or `$-$'; for example $f_{\rm 12+}$ = $f_{\rm 12}$ $+$ 1~d$^{-1}$ and
$f_{\rm 12-}$ = $f_{\rm 12}$ $-$ 1~d$^{-1}$. Table 2 of Bar05 indicated two possible 
aliases for a single frequency designated as $f_{\rm 18}$. We shall denote the
two possibilities as $f_{18^\prime}$ and $f_{18^{\prime\prime}}$, the frequency of the
former being lower than the latter. According to this notation, the frequency listed as
$f_{\rm 14}$ in Table 2 of \citet{baran08} should be designated as $f_{18^{\prime\prime}}$.


\subsection{Fourier analysis of the 2005 campaign data}
\label{ft}
\begin{figure}
\includegraphics[width=83mm]{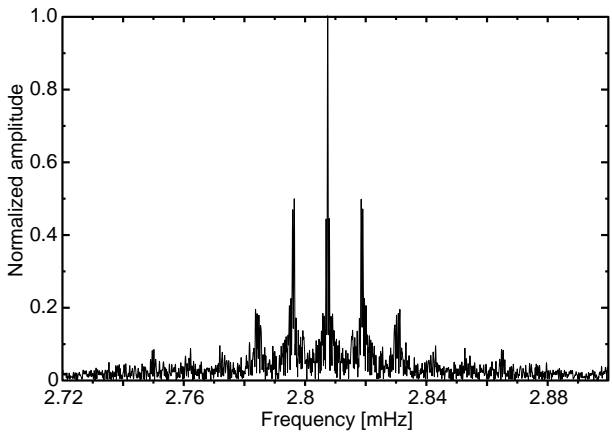}
\caption{Spectral window of 2005 $Bvb$ data of Bal\,09 shifted to the frequency of the strongest mode.}
\label{win_b_2005}
\end{figure}

Both the 2004 and 2005 data for Bal09 cover similar time intervals: nearly 40 days in 2004 and 50
days in 2005. The resulting resolution is therefore similar, but the improved duty cycle
of the 2005 data results in a lower detection threshold and improved aliasing. 
In terms of time coverage, the 2005 $UBVR$ and Str\"omgren $uvby$ data
are complementary.  Therefore, we decided to combine them. In order to minimize the wavelength dependence of
pulsation amplitudes, we combined the $U$ data with the mean of the $u$ and $v$ (hereafter $Uuv$) data, the
$B$ data with the average of the $v$ and $b$ (hereafter $Bvb$) data, and the $V$ and $y$ (hereafter $Vy$) data. 
{The $Bvb$ data are the most numerous of the four data sets and so provide the best spectral window which} is shown in Fig.~\ref{win_b_2005}.

\begin{figure}
\includegraphics[width=83mm]{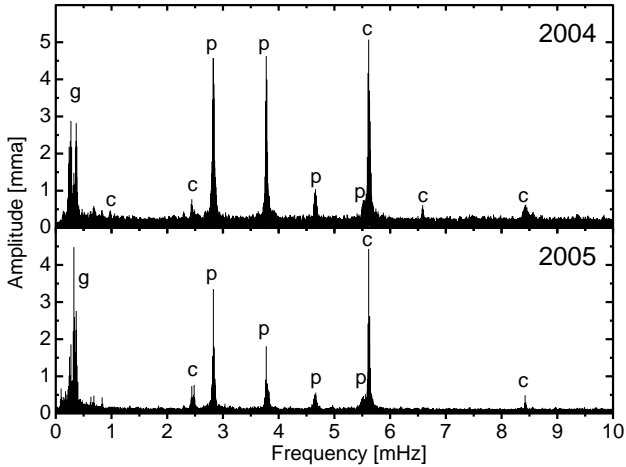}
\caption{Fourier spectrum of the 2004 $B$ (top) and 2005 $Bvb$ data (bottom) after removing the
three strongest frequencies near 2.8\,mHz ($f_1$, $f_2$ and $f_3$, see Tables \ref{a_fi_2004} and \ref{a_fi_2005}).
Regions attributed to $p$-modes, $g$-modes and their combinations are labeled with `p', `g', and `c', respectively.}
\label{ft_raw_b_2005}
\end{figure}

In Fig.~\ref{ft_raw_b_2005}, the Fourier spectrum for the $Bvb$ data is shown. Since oscillations in Bal\,09
are dominated by frequencies at 2.8~mHz, especially the strongest one with an amplitude near 50~mma, 
for clarity Fig.~\ref{ft_raw_b_2005} presents the amplitude spectrum after removing the three
highest-amplitude frequencies ($f_1$, $f_2$ and $f_3$).
It can be seen that the same groups of frequencies are present that were in the 2004 campaign (Bar05, Ore05): 
the low-frequency region (mainly around 0.3~mHz), that of the dominant frequency (at 2.8\,mHz) and
three other groups (3.8, 4.7, 5.5~mHz) separated roughly by the same frequency distance (0.8--1.0 mHz).
However, their amplitudes have changed; the low-frequency pulsations, in particular,
increased their amplitudes between 2004 and 2005.

\begin{figure}
\includegraphics[width=83mm]{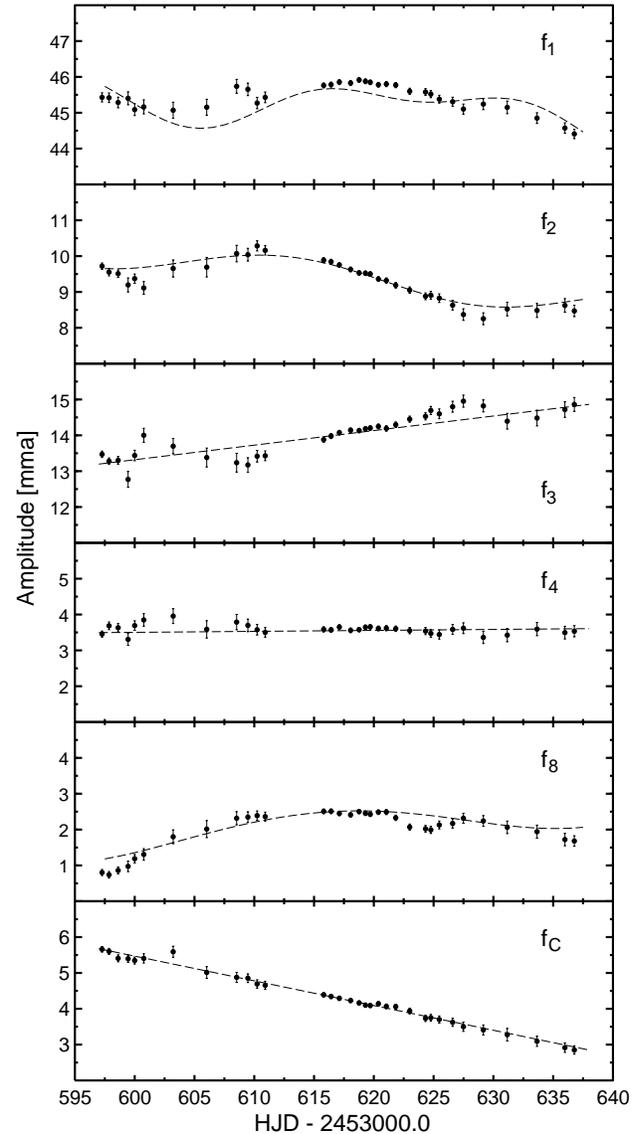}
\caption{Amplitudes of the six strongest frequencies in Bal\,09 for the 2005 $Bvb$ data. 
>From top to bottom the frequencies are $f_1$, $f_2$, $f_3$, $f_4$, $f_8$ and $f_{\rm C}$
(see Table \ref{a_fi_2005}). For $f_3$, $f_4$ and $f_{\rm C}$ the dashed lines correspond to changes
calculated from the derived values of $A$ and d$A$/d$t$ (Table \ref{a_fi_2005}), for the remaining three frequencies
they come from a more complicated model (see text for an explanation).}
\label{ampl_var_b_2005}
\end{figure}

>From the 2004 data Bar05 found only a weak evidence for amplitude variations, but 
those data were only sensitive to variations on a time scale of one or two
weeks. However, amplitude variations are obvious between the 2004 and 2005 data sets.
In order to infer the character of amplitude changes, we divided the 2005 $Bvb$ data into
33 subsets. The subsets contained data covering time intervals of 10 days in order to ensure that
the components of multiplets at 2.8\,mHz will be resolved. The subsets, however, largely overlap. The first subset started
at the beginning of $Bvb$ data, the next one a night later than the first one and so on.  The subsets were analysed
independently and the amplitudes for only the six strongest frequencies ($f_1$ through $f_4$, $f_8$, and $f_{\rm C}$) were used.
They are plotted in Fig.~\ref{ampl_var_b_2005}.  Of the six frequecies shown in Fig.~\ref{ampl_var_b_2005}, 
only $f_4$ shows no clear evidence for amplitude variations.

All four ($Uvb$, $Bvb$, $Vy$ and $R$) data sets were analysed independently.  
The procedure of extraction of periodic terms was a typical pre-whitening procedure, consisting of the following steps: 
(i) finding the frequency of the most significant peak in the Fourier periodogram calculated using the residuals from previous fit (or
the original light curve when starting the procedure), 
(ii) fitting a truncated Fourier series by means of least-squares with frequencies fixed; all previously found terms,
also the combination ones, were included in the fit,
(iii) improving the frequencies, phases and amplitudes of these terms using non-linear least-squares fit. The residuals from such
a fit were used in step (i). The steps (i)--(iii) were repeated until no significant peaks in the Fourier periodogram were found.

In view of significant amplitude variations observed in Bal\,09, in the final analysis we decided to allow for amplitude changes,
at least for the strongest modes.  Therefore, for the strongest terms the rates of amplitude changes were added 
in the step (iii) as free parameters, with d$A$/d$t$ = 0 as starting values. 
By the strongest terms we mean those which had amplitudes at least $\sim$3.5 times larger than the 
detection threshold in a given data set, i.e.~2.0, 0.7, 1.6 and 1.5\,mma, for $Uuv$, $Bvb$, $Vy$ and $R$ data, respectively.
(If for a given term the rate of amplitude change was derived, the resulting value is reported in Table \ref{a_fi_2005}.)
For the remaining terms, d$A$/d$t$ was fixed to zero. Fixing the rates to zero for the small-amplitude terms \
allowed us to avoid problems with convergence which otherwise occurred.
For $Uuv$, $Vy$ and $R$, the process was continued until all terms with signal-to-noise (S/N) above 4 were extracted. The resulting frequencies,
phases, amplitudes and rates of amplitude change are given in Table \ref{a_fi_2005}.

\begin{center}
\footnotesize
\begin{table*}
\caption{Modes detected in 2005 data in the $Uuv$, $Bvb$, $Vy$ and $R$ data. Numbers in parentheses indicate the r.m.s.~errors. Phases
and amplitudes are given for epoch HJD\,2\,453\,615.}
\begin{tabular}{clcccccccc}
\hline
  \# & Mean freq. & \multicolumn{4}{c}{Amplitude, $A$ [mma], d$A$\,/\,d$t$ [mma/d]} & \multicolumn{4}{c}{Phase [rad]} \\
    &      [mHz]       & $Uuv$ & $Bvb$ & $Vy$ & $R$ & $Uuv$ & $Bvb$ & $Vy$ & $R$ \\
\hline
$f_{\rm O}$ & 0.096902(09) & ...           & 0.70,$-$0.031 & ...           & ...           & ...      & 4.86(07) & ...      &          \\
$f_{\rm P}$ & 0.139161(22) & ...           &  ...          & 0.77, ---     & ...           & ...      & ...      & 1.15(13) &          \\
$f_{\rm Q}$ & 0.142459(10) & ...           & 0.59, ---     & ...           & ...           & ...      & 3.37(07) & ...      &          \\
$f_{\rm R}$ & 0.144996(22) & ...           &  ...          & 0.79, ---     & ...           & ...      & ...      & 5.65(13) &          \\
$f_{\rm S}$ & 0.169849(10) & ...           & 0.58, ---     & ...           & ...           & ...      & 4.09(07) & ...      &          \\
$f_{\rm T}$ & 0.175656(08) & ...           & 0.68, ---     & 0.88, ---     & ...           & ...      & 0.90(06) & 0.56(12) &          \\
$f_{\rm U}$ & 0.192869(21) & ...           &  ...          & 0.83, ---     & ...           & ...      & ...      & 3.41(12) &          \\
$f_{\rm V}$ & 0.198935(11) & ...           & 0.54, ---     & ...           & ...           & ...      & 5.89(08) & ...      &          \\
$f_{\rm I}$ & 0.229549(09) & ...           & 0.85,$-$0.027 & ...           & 0.49, ---     & ...      & 5.01(06) & ...      & 4.95(19) \\
$f_{\rm D}$ & 0.239971(07) & 1.68, ---     & 1.08,$+$0.019 & 0.99, ---     & 0.85, ---     & 5.81(08) & 5.99(04) & 5.98(10) & 5.73(11) \\
$f_{\rm F}$ & 0.246301(05) & 1.80, ---     & 1.52,$+$0.018 & 1.66,$-$0.004 & 1.37, ---     & 1.79(07) & 1.613(27)& 1.58(06) & 1.57(07) \\
$f_{\rm A}$ & 0.272463(05) & 1.67, ---     & 1.72,$-$0.002 & 1.45, ---     & 1.49, ---     & 1.37(08) & 1.473(23)& 1.58(07) & 1.57(07) \\
$f_{\rm W}$ & 0.29380(03)  & ...           & ...           & ...           & 0.52, ---     & ...      & ...      & ...      & 2.20(18) \\
$f_{\rm G+}$ & 0.310575(11) & 0.85, ---     & 0.72,$-$0.008 & 0.56, ---     & 0.58, ---     & 4.57(15) & 4.66(06) & 4.70(18) & 4.88(16) \\
$f_{\rm C}$ & 0.3256089(18)& 4.84,$-$0.077 & 4.44,$-$0.069 & 3.82,$-$0.019 & 3.71,$-$0.053 & 4.590(27)& 4.609(10)& 4.609(27)& 4.613(26)\\
$f_{\rm J}$ & 0.331182(10) & 0.71, ---     & 0.75,$-$0.002 & 0.79, ---     & 0.61, ---     & 0.87(18) & 6.19(05) & 0.58(13) & 6.25(15) \\
$f_{\rm B}$ & 0.365805(03) & 3.36,$-$0.014 & 2.65,$-$0.013 & 2.52,$-$0.006 & 2.64,$+$0.002 & 6.20(04) & 6.180(15)& 0.06(04) & 6.22(04) \\
$f_{\rm K}$ & 0.397232(09) & 0.67, ---     & 0.84,~~~0.000 & 0.97, ---     & 0.56, ---     & 3.01(19) & 2.52(05) & 2.63(11) & 2.52(17) \\
$f_{\rm L}$ & 0.630740(15) & ...           & 0.42, ---     & 0.49, ---     & ...           & ...      & 1.36(09) & 1.64(21) & ...      \\
$f_{\rm M}$ & 0.684405(14) & ...           & 0.47, ---     & 0.56, ---     & 0.38, ---     & ...      & 5.02(08) & 4.90(18) & 4.91(25) \\
$f_{\rm N}$ & 0.833090(14) & 0.59, ---     & 0.43, ---     & ...           & ...           & 4.67(22) & 4.77(09) & ...      & ...      \\
$f_{\rm X}$ & 1.845938(27) & ...           & 0.20, ---     & ...           & ...           & ...      & 4.48(19) & ...      & ...      \\
$f_{\rm Y}$ & 1.95968(04)  & 0.59, ---     & ...           & ...           & ...           & 0.55(22) & ...      & ...      & ...      \\
$f_{\rm Z}$ & 2.296144(22) & ...           & 0.25, ---     & ...           & ...           & ...      & 4.23(15) & ...      & ...      \\
\hline
$f_{\rm 1}-f_{\rm K}$ & 2.410232  & ...       & 0.22, ---     & ...       & ...       & ...      & 3.68(18) & ...      & ...      \\
$f_{\rm 1}-f_{\rm B}$ & 2.441659  & 0.86, --- & 0.69, ---     & 0.51, --- & 0.47, --- & 6.06(14) & 0.14(06) & 6.23(18) & 6.14(18) \\
$f_{\rm 1}-f_{\rm C}$ & 2.4818552 & 0.81, --- & 0.77,$-$0.013 & 0.93, --- & 0.63, --- & 1.81(14) & 1.98(05) & 1.74(10) & 1.78(14) \\
$f_{\rm 3}-f_{\rm C}$ & 2.4992272 & ...       & 0.20, ---     & ...       & ...       & ...      & 4.75(20) & ...      & ...      \\
\hline
$f_{\rm AA}$ & 2.707505(27) & ... & 0.20, --- & ... & ... & ... & 3.34(19) & ... & ... \\
\hline
2$f_{\rm 1}-f_{\rm 3}$& 2.7900921 & ... & 0.20, --- & ... & ... & ... & 1.31(19) & ... & ... \\
\hline
$f_{\rm 1}$ & 2.80746412(16)& 59.89,$-$0.048 & 45.48,$-$0.014 & 41.26,$+$0.019 & 40.08,$-$0.019 & 5.912(02)  & 5.934(01) & 5.927(03) & 5.934(03) \\
$f_{\rm 2}$ & 2.8230026(08) & 11.16,$-$0.013 &  9.49,$-$0.033 &  8.61,$-$0.012 &  8.27,$-$0.019 & 0.638(12)  & 0.658(05) & 0.653(12) & 0.626(12) \\
$f_{\rm 3}$ & 2.8248361(06) & 17.33,$+$0.029 & 13.93,$+$0.041 & 13.49,$+$0.015 & 12.88,$+$0.042 & 2.441(08)  & 2.457(03) & 2.454(08) & 2.428(08) \\
$f_{\rm 4}$ & 2.8265995(22) &  4.11,$+$0.035 &  3.54,$+$0.003 &  3.16,$+$0.001 &  2.99,$+$0.005 & 2.16(03)   & 2.095(12) & 2.01(04)  & 2.11(03) \\
$f_{\rm 5}$ & 2.853383(09)  &  1.06, ---     &  0.92,$-$0.011 &  0.58, ---     &  0.54, ---     & 1.00(12)   & 0.95(05)  & 0.61(18)  & 1.13(18) \\
$f_{\rm 7}$ & 2.855571(10)  &  0.66, ---     &  0.85,$-$0.008 &  0.82, ---     &  0.68, ---     & 5.16(20)   & 4.99(05)  & 5.00(13)  & 4.91(14) \\
$f_{\rm 36}$& 2.857126(13)  &  1.05, ---     &  0.59, ---     &  0.69, ---     &  0.59, ---     & 0.33(13)   & 0.54(07)  & 0.73(15)  & 0.80(17) \\
$f_{\rm 6}$ & 2.858725(10)  &  0.91, ---     &  0.88,$-$0.001 &  0.76, ---     &  0.96, ---     & 5.87(14)   & 5.74(05)  & 5.87(14)  & 5.67(10) \\
$f_{\rm 37}$& 2.860737(16)  &  ...           &  0.38, ---     &  ...           &  ...           & ...        & 4.94(11)  & ...       & ...      \\
\hline
$f_{\rm 38}$& 3.036271(21)  & ... &  0.28, --- & ... & ... & ... & 4.96(13) & ... & ... \\
\hline
$f_{\rm 1}+f_{\rm C}$ & 3.1330730 & ... & 0.22, --- & ... & ... & ... & 0.51(18) & ... & ... \\
$f_{\rm 1}+f_{\rm B}$ & 3.173269  & ... & 0.18, --- & ... & ... & ... & 2.36(22) & ... & ... \\
\hline
\end{tabular}
\label{a_fi_2005}
\end{table*}
\end{center}

\addtocounter{table}{-1}
\begin{center}
\footnotesize
\begin{table*}
\caption{continued.}
\begin{tabular}{clrrrrcccc}
\hline
\hline
  \# & Mean freq. & \multicolumn{4}{c}{Amplitude, $A$ [mma], d$A$\,/\,d$t$ [mma/d]} & \multicolumn{4}{c}{Phase [rad]} \\
    &      [mHz]       & $Uuv$ & $Bvb$ & $Vy$ & $R$ & $Uuv$ & $Bvb$ & $Vy$ & $R$ \\
\hline
$f_{\rm 39}$ & 3.774190(15) & 0.84, ---     &  0.49, ---     &  0.64, ---     & ...           & 4.81(15) & 3.96(09) & 3.92(16) & ...      \\
$f_{\rm 8}$  & 3.776163(05) & 2.30,$-$0.006 &  2.00,$+$0.023 &  2.26,$-$0.016 & 1.98,$+$0.008 & 4.04(06) & 4.03(03) & 3.95(05) & 4.02(05) \\
$f_{\rm 40}$ & 3.777480(16) & ...           &  0.46, ---     &  ...           & ...           & ...      & 1.49(10) & ...      & ...      \\
$f_{\rm 41}$ & 3.78451(03)  & ...           &  ...           &  0.52, ---     & ...           & ...      & ...      & 5.35(20) & ...      \\
$f_{\rm 42}$ & 3.784938(19) & ...           &  0.37, ---     &  ...           & ...           & ...      & 2.71(11) & ...      & ...      \\
$f_{\rm 43}$ & 3.78533(03)  & ...           &  ...           &  ...           & 0.55, ---     & ...      & ...      & ...      & 2.25(18) \\
$f_{\rm 44} = f_{\rm 9}$? & 3.786291(17) & ...           &  0.48, ---     &  ...           & 0.43, ---     & ...      & 4.14(09) & ...      & 4.21(23) \\
$f_{\rm 45} = f_{\rm 9}$? & 3.787153(21) & ...           &  0.37, ---     &  ...           & ...           & ...      & 4.18(14) & ...      & ...      \\
$f_{\rm 46}$ & 3.791291(15) & ...           &  0.58, ---     &  ...           & 0.75, ---     & ...      & 2.70(08) & ...      & 2.47(14) \\
$f_{\rm 24}$ & 3.791581(12) & 1.28, ---     &  0.78,$+$0.030 &  0.74, ---     & 0.57, ---     & 0.91(10) & 0.76(08) & 1.10(14) & 0.56(19) \\
$f_{\rm 47}$ & 3.792387(22) & ...           &  0.36, ---     &  ...           & ...           & ...      & 0.62(13) & ...      & ...      \\
$f_{\rm 48}$ & 3.793419(20) & ...           &  0.35, ---     &  ...           & ...           & ...      & 4.28(12) & ...      & ...      \\
$f_{\rm 49}$ & 3.796825(20) & 0.61, ---     &  0.39, ---     &  ...           & ...           & 5.45(21) & 5.79(12) & ...      & ...      \\
$f_{\rm 19}$ & 3.80520(04)  & ...           &  ...           &  0.51, ---     & ...           & ...      & ...      & 1.09(20) & ...      \\
$f_{\rm 20}$ & 3.806574(22) & ...           &  0.30, ---     &  ...           & ...           & ...      & 0.86(14) & ...      & ...      \\
$f_{\rm 50}$ & 3.807605(19) & ...           &  0.36, ---     &  ...           & ...           & ...      & 5.11(12) & ...      & ...      \\
$f_{\rm 18^{\prime\prime}}$ & 3.808945(15) & ...          &  0.44, ---     &  ...           & ...           & ...      & 4.45(09) & ...      & ...      \\
$f_{\rm 51}$ & 3.812194(17) & ...           &  0.37, ---     &  ...           & ...           & ...      & 0.97(11) & ...      & ...      \\
$f_{\rm 52} = f_{\rm 26}$? & 3.822485(17)   & ...            &  0.34, ---     &  ...          & ...      & ...      & 5.13(12) & ...      & ...      \\
$f_{\rm 53}$ & 3.824759(12) & ...           &  0.55, ---     &  ...           & 0.48, ---     & ...      & 1.71(07) & ...      & 1.92(20) \\
$f_{\rm 54}$ & 3.829725(16) & ...           &  0.42, ---     &  ...           & ...           & ...      & 2.17(12) & ...      & ...      \\
$f_{\rm 55}$ & 3.840824(23) & ...           &  0.28, ---     &  ...           & ...           & ...      & 3.15(17) & ...      & ...      \\
\hline
$f_{\rm 56}$ & 4.636265(26) & ... &  0.25, --- &  ...       &  ...       & ... & 3.45(17) & ...      & ...      \\
$f_{\rm 57}$ & 4.639892(15) & ... &  0.42, --- &  ...       &  ...       & ... & 3.87(10) & ...      & ...      \\
$f_{\rm 58}$ & 4.641239(23) & ... &  0.36, --- &  ...       &  0.42, --- & ... & 3.37(12) & ...      & 3.87(22) \\
$f_{\rm 59}$ & 4.641568(19) & ... &  0.47, --- &  0.59, --- &  ...       & ... & 3.17(10) & 2.93(17) & ...      \\
$f_{\rm 60}$ & 4.642478(16) & ...       &  0.49, --- &  ...       & ...        & ...      & 2.91(10) & ...      & ...      \\
$f_{\rm 61} = f_{\rm 11}$? & 4.645175(28) & ...       &  0.27, --- &  ...       & ...        & ...      & 1.59(17) & ...      & ...      \\
$f_{\rm 62} = f_{\rm 11}$? & 4.645459(28) & ...       &  0.33, --- &  ...       & ...        & ...      & 5.54(13) & ...      & ...      \\
$f_{\rm 63}$ & 4.646299(16) & ...       &  0.54, --- &  ...       & ...        & ...      & 4.73(08) & ...      & ...      \\
$f_{\rm 64}$ & 4.647062(28) & ...       &  0.28, --- &  ...       & ...        & ...      & 1.77(18) & ...      & ...      \\
$f_{\rm 65}$ & 4.649385(22) & 0.64, --- &  0.29, --- &  ...       & ...        & 0.11(20) & 0.16(15) & ...      & ...      \\
$f_{\rm 66}$ & 4.65294(04)  & 0.66, --- &  ...       &  ...       & ...        & 0.72(20) & ...      & ...      & ...      \\
$f_{\rm 67}$ & 4.653491(18) & ...       &  0.44, --- &  ...       & ...        & ...      & 4.11(12) & ...      & ...      \\
$f_{\rm 68}$ & 4.654759(20) & 0.74, --- &  0.34, --- &  ...       & ...        & 4.14(18) & 3.71(13) & ...      & ...      \\
$f_{\rm 12-}$ & 4.658007(19) & ...       &  0.41, --- &  ...       & ...        & ...      & 4.32(12) & ...      & ...      \\
$f_{\rm 69} = f_{\rm 13}$?& 4.66097(03)  & ...       &  ...       &  0.53, --- & ...        & ...      & ...      & 4.14(19) & ...      \\
$f_{\rm 70} = f_{\rm 13}$?& 4.661708(15) & ...       &  0.40, --- &  ...       &  ...       & ...      & 2.02(10) & ...      & ...      \\
$f_{\rm 71}$ & 4.663070(25) & ...       &  ...       &  ...       &  0.62, --- & ...      & ...      & ...      & 3.36(15) \\
$f_{\rm 72}$ & 4.665422(15) & ...       &  0.55, --- &  0.46, --- &  ...       & ...      & 4.36(08) & 4.58(23) & ...      \\
$f_{\rm 73}$ & 4.666868(15) & 0.64, --- &  0.58, --- &  0.37, --- &  0.55, --- & 5.96(20) & 6.01(07) & 6.10(28) & 6.25(17) \\
$f_{\rm 74}$ & 4.668308(18) & ...       &  0.37, --- &  ...       &  ...       & ...      & 4.22(11) & ...      & ...      \\
$f_{\rm 75}$ & 4.671570(23) & ...       &  0.29, --- &  ...       &  ...       & ...      & 0.75(15) & ...      & ...      \\
$f_{\rm 76}$ & 4.674040(26) & ...       &  0.24, --- &  ...       &  ...       & ...      & 1.98(17) & ...      & ...      \\
$f_{\rm 77}$ & 4.683263(23) & ...       &  0.28, --- &  ...       &  ...       & ...      & 0.57(15) & ...      & ...      \\
$f_{\rm 78}$ & 4.708165(26) & ...       &  0.21, --- &  ...       &  ...       & ...      & 0.25(18) & ...      & ...      \\
\hline
\end{tabular}
\end{table*}
\end{center}

\addtocounter{table}{-1}
\begin{center}
\footnotesize
\begin{table*}
\caption{continued.}
\begin{tabular}{clrrrrcccc}
\hline
\hline
  \# & Mean freq. & \multicolumn{4}{c}{Amplitude, $A$ [mma], d$A$\,/\,d$t$ [mma/d]} & \multicolumn{4}{c}{Phase [rad]} \\
    &      [mHz]       & $Uuv$ & $Bvb$ & $Vy$ & $R$ & $Uuv$ & $Bvb$ & $Vy$ & $R$ \\
\hline
$f_{\rm 79}$ & 5.468920(22) & ... &  0.26, --- &  ...  &  ...       & ... & 2.82(15) & ... & ...      \\
$f_{\rm 80}$ & 5.476806(23) & ... &  0.25, --- &  ...  &  ...       & ... & 4.63(16) & ... & ...      \\
$f_{\rm 81}$ & 5.477799(25) & ... &  0.24, --- &  ...  &  ...       & ... & 2.36(16) & ... & ...      \\
$f_{\rm 82}$ & 5.484197(24) & ... &  0.25, --- &  ...  &  ...       & ... & 5.66(15) & ... & ...      \\
$f_{\rm 83}$ & 5.487352(19) & ... &  0.32, --- &  ...  &  ...       & ... & 2.58(13) & ... & ...      \\
$f_{\rm 84}$ & 5.491241(16) & ... &  0.39, --- &  ...  &  ...       & ... & 2.90(11) & ... & ...      \\
$f_{\rm 85}$ & 5.496090(20) & ... &  0.30, --- &  ...  &  0.46, --- & ... & 0.77(13) & ... & 1.24(20) \\
$f_{\rm 86}$ & 5.503194(26) & ... &  0.24, --- &  ...  &  ...       & ... & 5.63(17) & ... & ...      \\
$f_{\rm 87}$ & 5.510788(25) & ... &  0.24, --- &  ...  &  ...       & ... & 4.45(17) & ... & ...      \\
$f_{\rm 88}$ & 5.516547(19) & ... &  0.30, --- &  ...  &  ...       & ... & 2.13(13) & ... & ...      \\
$f_{\rm 32^\prime}$ & 5.517516(25) & ... &  0.23, --- &  ...  &  ...       & ... & 0.11(17) & ... & ...      \\
$f_{\rm 89}$ & 5.522630(22) & ... &  0.28, --- &  ...  &  ...       & ... & 5.07(15) & ... & ...      \\
$f_{\rm 90}$ & 5.531191(15) & ... &  0.38, --- &  ...  &  ...       & ... & 3.81(10) & ... & ...      \\
$f_{\rm 91}$ & 5.534807(22) & ... &  0.27, --- &  ...  &  ...       & ... & 0.56(15) & ... & ...      \\
$f_{\rm 92}$ & 5.544758(22) & ... &  0.26, --- &  ...  &  ...       & ... & 3.58(15) & ... & ...      \\
$f_{\rm 93}$ & 5.593452(26) & ... &  0.21, --- &  ...  &  ...       & ... & 3.41(18) & ... & ...      \\
\hline
2$f_{\rm 1}$         & 5.61492824& 4.92,$+$0.025 & 4.39,$-$0.008 & 3.89,$+$0.008 & 3.63,$-$0.002 & 4.025(23) & 4.088(09) & 4.124(24)& 4.092(023) \\
$f_{\rm 1}+f_{\rm 2}$& 5.6304667 & 1.83, ---     & 1.79,$-$0.019 & 1.69,$-$0.008 & 1.30, ---     & 5.12(07)  & 5.089(22) & 5.19(06) & 5.13(07) \\
$f_{\rm 1}+f_{\rm 3}$& 5.6323002 & 2.99, ---     & 2.58,$+$0.011 & 2.23,$+$0.023 & 2.34,$-$0.013 & 0.67(04)  & 0.647(16) & 0.63(04) & 0.62(04) \\
$f_{\rm 1}+f_{\rm 4}$& 5.6340636 & ...           & 0.69, ---     & 0.89, ---     & 0.52, ---     & ...       & 0.25(06)  & 0.36(11) & 0.49(17) \\
$f_{\rm 2}+f_{\rm 3}$& 5.6478387 & ...           & 0.23, ---     & ...           & ...           & ...       & 1.84(17)  & ...      &          \\
2$f_{\rm 3}$         & 5.6496722 & ...           & 0.33, ---     & ...           & ...           & ...       & 2.85(12)  & ...      &          \\
$f_{\rm 1}+f_{\rm 6}$& 5.666189  & ...           & 0.18, ---     & ...           & ...           & ...       & 3.65(22)  & ...      &          \\
\hline
3$f_{\rm 1}$          & 8.4223924 & ... & 0.49, ---     & ...  & 0.60, --- & ... & 2.57(08) & ... & 2.63(14) \\
2$f_{\rm 1}+f_{\rm 2}$& 8.4379308 & ... & 0.21, ---     & ...  & ...       & ... & 3.44(19) & ... & ...      \\
2$f_{\rm 1}+f_{\rm 3}$& 8.4397643 & ... & 0.22, ---     & ...  & ...       & ... & 5.31(18) & ... & ...      \\
\hline
\multicolumn{2}{r}{N$_{\rm obs}$}            & 21251 & 63497 & 17307 & 18375 & & & & \\
\multicolumn{2}{r}{$\sigma_{A}$ [mma]}       & 0.12  & 0.04  & 0.09  & 0.09  & & & & \\
\multicolumn{2}{r}{$\sigma_{dA/dt}$ [mma/d]} & 0.012 & 0.003 & 0.010 & 0.009 & & & & \\
\multicolumn{2}{r}{Detection threshold [mma]}& 0.58  & 0.20  & 0.46  & 0.42  & & & & \\
\multicolumn{2}{r}{Residual SD [mma]}        & 12.05 & 6.59  & 8.58  & 8.08  & & & & \\
\end{tabular}
\end{table*}
\end{center}

For the $Bvb$ data, we used the same detection
limit (S/N $>$ 4), but the process required a modification
as the non-linear least-squares routines did not reach convergence once a large number of frequencies were fit.
Therefore, we first obtained a preliminary solution which included all frequencies with amplitudes higher
than 1\,mma. Next, we took advantage of the fact
that in Bal\,09 frequencies appear in groups separated by several hundred $\mu$Hz. Thus, we could safely assume that
during least-squares fits, frequencies from one group do not affect the solution for frequencies of the other groups. 
We obtained the final solution for $Bvb$ data using six groups. In this process, we first subtracted frequencies
from the other groups taken from the preliminary solution and  used the residuals as a starting point for the 
extraction of frequencies in the considered group.

As can be seen from Fig.~\ref{ampl_var_b_2005}, the amplitude variations can be more complicated than linear.
We may therefore expect the occurence of residual signals in the vicinity of strong terms resulting from amplitude
changes unaccounted for by the model we use, i.e., linear amplitude change. We detected several such terms (we call them `ghost' terms)
near $f_1$, $f_2$ and $f_8$. Since they are artefacts caused by amplitude changes not accounted for by a constant d$A$/d$t$,
they are not listed in Table \ref{a_fi_2005}. Having their amplitudes, frequencies and phases, we can, however, combine them with
the main term to show the {\it apparent} amplitude changes. This is shown 
in Fig.~\ref{ampl_var_b_2005} for $f_1$, $f_2$ and $f_8$. For the remaining terms shown in this figure the `ghost' terms were not
detected which means that the model with constant rate of amplitude change is good enough. Indeed, there are no large discrepancies from 
linear amplitude change for $f_3$, $f_4$ and $f_{\rm C}$.

The results of frequency extraction are presented in the following subsections
and the full list is given in Table \ref{a_fi_2005}. The largest number
of frequencies, 103, was detected in the $Bvb$ data, as expected. In each of the remaining three datasets
over 30 frequencies were detected. The total number of frequencies presented in Table \ref{a_fi_2005} is
114, since some of them which were not detected in $Bvb$ data were found in other datasets.
Of these 114 frequencies, 17 are combination terms while the remaining 97 may represent independent modes.

As indicated by Fig.~\ref{ampl_var_b_2005}, the amplitudes of some terms change by a factor of two or more
during 50 days (e.g. $f_8$ and $f_{\rm C}$). As such, we cannot exclude
the possibility that some nearby frequencies given in Table \ref{a_fi_2005} are not independent but are
artefacts caused by amplitude variations. 
Another problem is aliasing. This is particularly bad near 3.8, 4.7 and 5.5\,mHz where low-amplitude frequencies
are near to high-amplitude ones and the frequency density is very high. Again, it is
possible that some of these frequencies are in error
by 1~(sidereal day)$^{-1}$.
Furthermore, some frequencies may also have changed. While these changes probably did not affect
the analysis of any given year data, season-to-season changes
are evident for some components of rotationally split multiplets, as discussed in Section \ref{pran1}.
The changes of frequencies between 2004 and 2005 could be as large as 0.6\,$\mu$Hz (see Table \ref{spl_change}), which may pose
difficulties in cross-identifying frequencies detected in both seasons. That is why we provide multiple identifications
and question marks for some of the frequencies in Table \ref{a_fi_2005}.

\subsection{Frequencies in the $g$-mode region}
>From the 2005 data, we recovered all 12 frequencies below 0.9\,mHz found in the combined 2004 data (Table \ref{a_fi_2004}).
The frequency $f_{\rm G+}$, an alias of $f_{\rm G}$, was detected in the 2005 data and we think that
it is the true frequency of this mode. In addition,
a dozen new frequencies in this region were found. We notice that many of the
amplitudes changed considerably between 2004 and 2005. While the 2004 spectrum was dominated by
$f_{\rm A}$ and $f_{\rm B}$ with $B$ amplitudes near 2.7\,mma, the 2005 spectrum is dominated by 
$f_{\rm C}$ with an amplitude changing from $\sim$6\,mma to $\sim$3\,mma (Fig.~\ref{ampl_var_b_2005}).
\begin{figure}
\includegraphics[width=83mm]{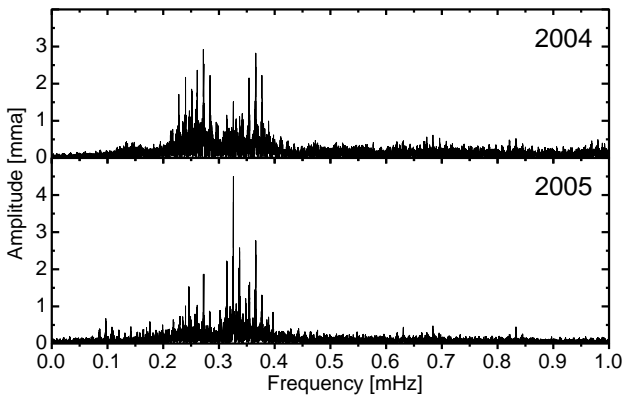}
\caption{\small Fourier spectrum of Bal09 below 1\,mHz for $B$ 2004 (top) and $Bvb$ 2005 (bottom) data.}
\label{spec_g}
\end{figure}

The frequencies which we interpret as $g$-modes occur mainly between 0.1 and 0.4\,mHz. However, several
periodicities at higher frequencies, starting from 0.6\,mHz up to $f_{\rm AA}$ at 2.7\,mHz, were found.  Some
of them were already detected in the 2004 data.  It is difficult to prove this, but the comparison with model frequencies 
indicates at least for those with the lowest frequencies, the most plausible explanation is also a $g$-mode pulsation.

\subsection{The region of the dominant frequency (2.8~mHz)}
\label{pran1}
\begin{figure}
\includegraphics[width=83mm]{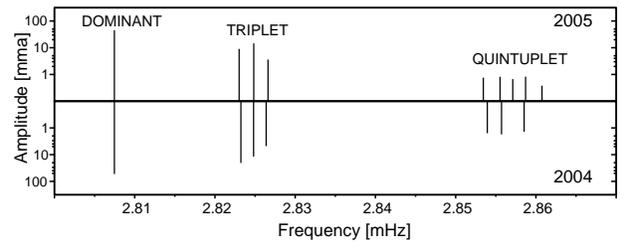}
\caption{\small Schematic amplitude spectra near 2.8~mHz of 2005 (top) and 2004 data (bottom).}
\label{schematic_ft_2004_2005}
\end{figure}

This is the first of four narrow regions in frequency where $p$-modes occur. The $f_1$ (the highest amplitude term) has been previously
identified as radial mode by \citet{baran08}. The $f_2$, $f_3$ and $f_4$ form an equidistant triplet, presumably a rotationally split
$\ell$ = 1 mode (Bar05) and we also recovered $f_5$, $f_6$ and $f_7$ in this region.
Additionally, two new frequencies,
$f_{\rm 36}$ and $f_{\rm 37}$, were found in the 2005 data. Together with $f_5$, $f_6$ and $f_7$ they form a nearly symmetric, although not
equally-spaced, quintuplet which is probably a $\ell$ = 2 rotationally split mode. This was partially confirmed by
\citet{baran08} using multicolour and spectroscopic observations. In addition to the two multiplets and the radial mode, we found no other
frequencies in the region of 2.8\,mHz except the `ghost' terms already discussed in Section \ref{ft}.
We also noticed that the multiplet spacings changed between 2004 and 2005, becoming wider with time.
This is shown schematically in Fig.~\ref{schematic_ft_2004_2005}. We will discuss the possible explanation of
this behaviour in Section \ref{concl}.

\begin{table}
\centering
\caption{Frequencies and splittings for all modes near 2.85\,mHz.}
\begin{tabular}{ccccc}
\hline
   &        & Frequency & Frequency & Difference \\
\# & $(\ell,m)$ & 2004 & 2005 & in freq. \\
   &     &   [mHz]   & [mHz] & [$\mu$Hz] \\
\hline
$f_{\rm 1}$ & (0,0)    & 2.8074697(02)& 2.80746412(16) & $-$0.0056(03) \\
\hline
$f_{\rm 2}$ & (1,$+$1) & 2.8232390(06)& 2.8230026(08) & $-$0.2364(10) \\
$f_{\rm 3}$ & (1,0)    & 2.8248080(11)& 2.8248361(06) & $+$0.0281(13) \\
$f_{\rm 4}$ & (1,$-$1) & 2.8263696(27)& 2.8265995(22) & $+$0.2315(35) \\
\hline
$f_{\rm 5}$ & (2,$+$2) & 2.853958(08) & 2.853383(09) & $-$0.575(12) \\
$f_{\rm 7}$ & (2,$+$1) & 2.855710(10) & 2.855571(10) & $-$0.139(14) \\
$f_{\rm 36}$ & (2,0)   & ---          & 2.857126(13) & --- \\
$f_{\rm 6}$ & (2,$-$1) & 2.858534(14) & 2.858725(10) & $+$0.191(17) \\
$f_{\rm 37}$ & (2,$-$2)& ---          & 2.860737(21) & --- \\
\hline
\multicolumn{2}{c}{} & Splitting & Splitting & Change\\
\multicolumn{2}{c}{Difference} & 2004 & 2005 & of splitting \\
\multicolumn{2}{c}{} & [$\mu$Hz] & [$\mu$Hz] & [\%]\\
\hline
\multicolumn{2}{c}{$f_{\rm 3}-f_{\rm 2}$} & 1.5690(13) & 1.8336(08) & $+$16.86(11) \\
\multicolumn{2}{c}{$f_{\rm 4}-f_{\rm 3}$} & 1.5616(29) & 1.7650(18) & $+$13.03(24) \\
\hline
\multicolumn{2}{c}{$f_{\rm 7}-f_{\rm 5}$}  & 1.752(13)   & 2.188(13) & $+$24.9(12) \\
\multicolumn{2}{c}{$f_{\rm 36}-f_{\rm 7}$} & ---         & 1.555(16) & ---        \\
\multicolumn{2}{c}{$f_{\rm 6}-f_{\rm 36}$} & ---         & 1.599(16) & ---        \\
\multicolumn{2}{c}{$f_{\rm 37}-f_{\rm 6}$} & ---         & 2.012(23) & ---        \\
\multicolumn{2}{c}{$f_{\rm 6}-f_{\rm 7}$}  & 2.824(17)   & 3.154(14) & $+$11.7(08) \\
\hline
\end{tabular}
\label{spl_change}
\end{table}

\begin{figure}
\includegraphics[width=83mm]{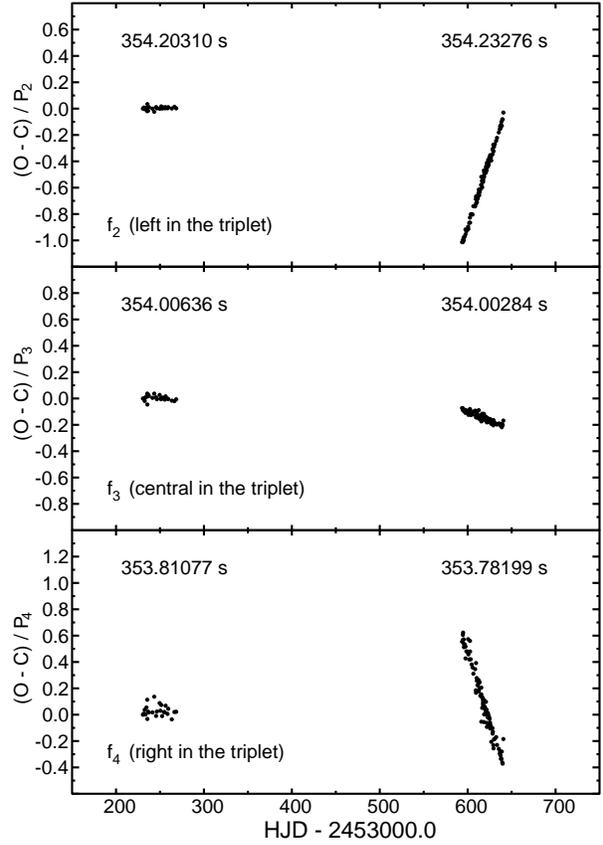}
\caption{\small The O$-$C diagrams calculated from the nightly times of maximum light for the components of the 2.82\,mHz triplet.
Periods corresponding to seasonal mean values are given in the panels.}
\label{oc2}
\end{figure}

The frequencies of the multiplet components and the splittings for 2004 and 2005 are
given in Table \ref{spl_change}. We can see from this table that while the components of the triplet were within the errors
equidistant in 2004, this is no longer the case in 2005 and a small asymmetry appeared.
Moreover, the splittings were larger by about 15\% in 2005. Not only the side components of the triplet
changed their frequencies, but that of the central one also changed by a significant amount.
The changes of frequencies of the triplet components can be even better illustrated in the O$-$C diagram
(Fig.~\ref{oc2}) where nightly times of maximum light were used. For each frequency, they were obtained by subtracting all other frequencies
from the $Bvb$ data and then dividing the residuals into separate nights. Then, times of maximum brightness were derived for
each night. It can be seen that while within the time interval
covered by observations in each season (40--50 days), the period can be regarded as nearly constant (though different for
2004 and 2005), the changes of periods between the two seasons are evident and very large for $f_2$ and $f_4$.
Unfortunately, it is not known how fast and when the periods underwent such a large change.
There is an obvious ambiguity in counting pulsation cycles between 2004 and 2005 data. This means that in Fig.~\ref{oc2}
the 2005 points could be shifted upward or downward by an integer number of pulsation cycles.

A comparison of the 2004 and 2005 frequencies of the components of the quintuplet indicates
that the quintuplet behaved in a similar way as the triplet, i.e., increased spacings.
The change in spacings was different for the $|m|$ = 1 and $|m|$ = 2 components,
but as a first approximation we can claim that it was proportional to $|m|$. We may therefore speculate that 
the same mechanism caused the change of splittings in both multiplets.
It is interesting to note that the main frequency, $f_1$, also changed slightly between 2004 and 2005. It would be interesting
to monitor this behaviour on a longer time scale as it could be used to determine evolutionary changes of the star \citep{reed04}.

Amplitudes have also changed for these frequencies.
The largest change was observed for $f_2$ which decreased from 20.4 to 9.5\,mma in $B$/$Bvb$ data, but
the amplitudes of $f_1$, $f_4$, $f_5$, $f_6$ and $f_7$ also decreased from 2004 to 2005. Only for $f_3$ did the amplitude
slightly increase.

\subsection{Other regions with $p$-modes}
There are three other frequency regions where $p$-modes occur, namely at 3.8\,mHz, 4.7\,mHz and 5.5\,mHz.
These regions contain frequencies with very small amplitudes. Most of them were detected only in the $Bvb$ data and
very few of these were found in the 2004 data. Of the frequencies from 2004 that were recovered in the 2005 data, they
typically had smaller amplitudes, $f_8$
being an example. A proper correlation of the 2004 frequencies to those in the 2005 data has
three complications:
(i) The presence of many frequencies with similarly small amplitudes; (ii) strong aliasing; and
(iii) frequencies corresponding to $|m| >$ 0 modes might have changed their frequencies considerably between 2004 and 2005, as
seen in the multiplets.
These uncertainties lead to ambiguities in the frequency correlation which are indicated in Table \ref{a_fi_2005} using
double identifications and question marks.

As the frequency densities are high within each region, some of the frequencies are likely members of rotationally split multiplets. 
However, $\ell$ and $m$ mode assignment would be uncertain from our data. Nevertheless, their frequencies
may be useful for asteroseismology once the character of long-term frequency-spacing changes is established. It is also possible that
other, currently undetected components of multiplets will be detected in future observations, allowing the full structure of
multiplets to be observed. These regions are therefore potentially very important for future asteroseismological analyses.

There is also an isolated frequency ($f_{38}$) near 3.036\,mHz, which is between $p$-mode regions. 
Its amplitude is very small but its detection has been confirmed by G.\,Fontaine (private communication) in an independent set of data,
and so we consider it to be intrinsic to Bal09.

\subsection{Combination frequencies}
>From the 2005 data, 17 combination terms were detected in five distinct regions. They are typically
a simple sum or difference of two frequencies. Of them, the most important are those that involve
$g$- and $p$-modes as it proves that both kinds of modes occur in the same star. They were previously detected by Bar05 and
are detected in the combined 2004 data. There are six $g$- and $p$-mode combination frequencies observed in the 2005 data and
three in the 2004 data.
We do not recover the $f_8 - f_1$ and $f_1 + f_8$ combinations present in the 
2004 data (Table \ref{a_fi_2004}), but we do detect combination frequencies in two new regions: sums of $g$- and
$p$-modes and $2f_1 - f_3$.
We also detect two harmonics of $f_1$. Their amplitudes and phases indicate that
despite a decrease of the $B$-filter amplitude by about 14\% between 2004 and 2005, the shape of the light curve of the main mode
remained practically unchanged.

\subsection{Remarks on the frequency spectrum}
Before we try to compare observed frequencies with those derived from stellar models, a few general comments on the
frequency spectrum need to be made. The substantial data sets of 2004 and 2005
indicate undoubtedly that amplitude variations are very common in this star. Some amplitudes changed
by a factor of  2--3 over the span of our data. Figure \ref{ampl_var_b_2005} indicates that the time scale of amplitude 
changes might be as short as two weeks.
Amplitude changes were also observed in other sdBV stars. For example, \cite{kilkenny06} present the results for four sdBV stars which
have been observed at least twice. Their analysis showed that for three of them the amplitudes of some modes changed significantly. 
For example, for EC\,20338$-$1925 the dominant mode decreased its amplitude from 26 to 4\,mmag. A similar change was found
for V\,338~Ser.  Amplitude changes have also been observed in other sdBV stars. \citet{reed07b} examined 
follow-up data on 23 sdBV stars (including some Bal09 data) and of 54 consistently detected frequencies, 34 (63\%)
changed amplitudes by more than a factor of two over the duration of their observations (from a week up to several years). 
Fourteen of their frequencies
changed amplitudes by more than a factor of five; including one that ranged from 57 to 8~mma.
Therefore, the amplitude variations  in Bal09 are consistent with those observed in many other sdBV stars.

We also detect a clear change of frequencies, especially of those known to have $|m|\ne$ 0, i.e., the side
components of the rotationally split multiplets. A small frequency change of the radial mode $f_1$ was also detected.
It would be interesting to trace these changes during a longer time scale, not only in Bal09, but also in other sdBV stars.
They can be used to constrain the evolution of sdB stars and/or indicate the mechanism that
causes these changes.

\section{Comparison with theoretical frequencies}
\subsection{The models}
\label{ev_models}%
\begin{figure}
\centering
\includegraphics[width=83mm]{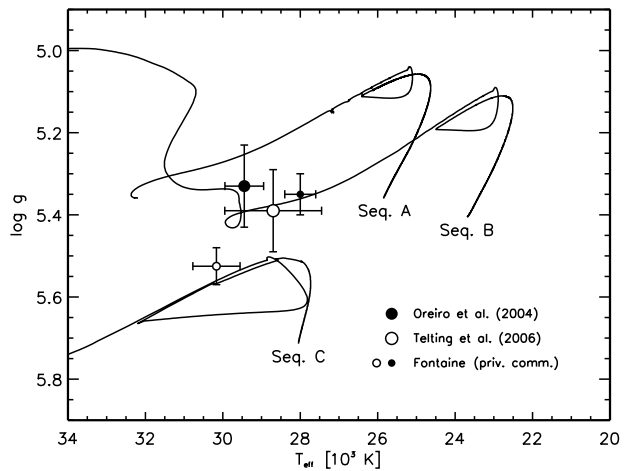}
\caption{Physical parameters for Bal\,09 derived by different authors. Three sdB evolutionary sequences
(A, B, C) used for theoretical comparison are also shown.  
He-core burning starts at the point closest to the sequence name for each
evolutionary track. The lowest gravity phase of a track corresponds to the
He-core content of $\sim$0.1, when the core abruptly engulfs material from the
radiative zone, causing the loops seen in the tracks at this evolutionary
stage.}
\label{dhr}%
\end{figure}

Once the observed frequencies have been determined, it is interesting to compare them with theoretical
frequencies computed for models of sdB stars. For this purpose, we computed three sdB evolutionary sequences (A, B, C).
These sequences begin at the onset of He-core burning and their parameters are given in Table \ref{models}, while the 
evolutionary tracks are shown in the $T_{\rm eff}$ -- $\log g$ diagram of Fig.~\ref{dhr}.  For
comparison, four spectroscopic determinations of $T_{\rm eff}$ and $\log g$ of Bal09 are also shown with error bars. 

Sequences B and C were obtained by evolving an 1\,$M_{\odot}$ star from the main sequence using the stellar
evolution code of~\cite{jimenez96}. A different mass loss Reimer's parameter on the red giant branch leads to sdB tracks
with different H-envelope mass ($M_{\rm H}$), although with very similar total mass ($\sim$0.47\,$M_{\odot}$),
as the core needed to ignite He-core burning is almost the same in all cases. No binary mass transfer was included.
The position of an evolutionary sequence in the $T_{\rm eff}-\log g$ plane depends mainly on the total mass
($M$) and on $M_{\rm H}$. Tracks with the same $M$ are located along a diagonal in this plane, ordered by $M_{\rm H}$: the lower
$M_{\rm H}$, the larger $T_{\rm eff}$ and $\log g$ are. The diagonal is shifted to higher (lower) temperatures if $M$
is increased (reduced). Thus, comparing the spectroscopic parameters of Bal09 with the location of sequences B and
C (Fig.~\ref{dhr}), we may conclude that either the star is at the final stages of He-core burning (or past it) or it is
at the beginning of the sdB phase, but with a total mass higher than 0.47\,$M_{\odot}$. 
In the latter case, formation channels involving binary interaction should be included to account for the higher mass.
To allow for the posibility of higher mass for Bal09, a third evolutionary sequence (A) was computed ad hoc with 
a slightly higher total mass (0.55\,$M_{\odot}$) than for sequences B and C and was evolved with the same
code. Its parameters 
at He-core ignition are also included in Table~\ref{models}.
\begin{table} %
\centering
\small
\caption[]{Parameters of the models at the beginning of the He-core burning phase for three
evolutionary sequences. Provided are the total mass ($M$), H-envelope mass ($M_{\rm H}$), effective temperature ($T_{\rm eff}$)
and logarithm of surface gravity ($\log g$).}
\begin{tabular}{lccccc}
\hline
Model  & $M$ & $M_{\rm H}$       & $T_{\rm eff}$ & $\log g$\\
       & [$M_{\odot}$]   &[$M_{\odot}$]      & [K]           & [dex]\\
\hline
A        & 0.554           & 2.9$\cdot$10$^{-3}$ & 25\,860 & 5.36\\
B        & 0.475           & 3.1$\cdot$10$^{-3}$ & 23\,690 & 5.40\\
C        & 0.473           & 4.6$\cdot$10$^{-4}$ & 28\,060 & 5.71\\
\hline
\end{tabular}
\label{models}
\end{table}

\subsection{Analysis of $p$-modes}
\begin{figure}
\centering
\includegraphics[width=83mm]{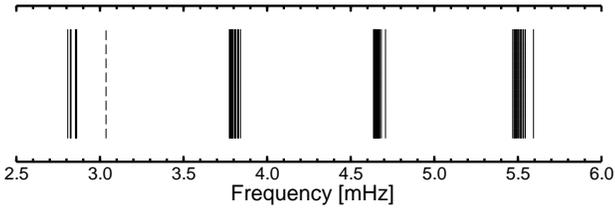}
\caption{Schematic frequency spectrum of Bal\,09 in the region where the likely $p$-modes occur.
The dashed line corresponds to $f_{38}$.}
\label{esquemap}%
\end{figure}

In Fig.~\ref{esquemap} a schematic spectrum of Bal\,09 in the $p$-mode domain is presented. 
As can be seen, all $p$-modes except $f_{38}$ cluster in four narrow (from 0.05 to 0.12 mHz wide) frequency
groups near 2.8, 3.8, 4.7 and 5.5 mHz, consisting of 9, 22, 24, and 16 frequencies, respectively (Table \ref{a_fi_2005}).

\begin{figure}
\centering
\includegraphics[width=83mm]{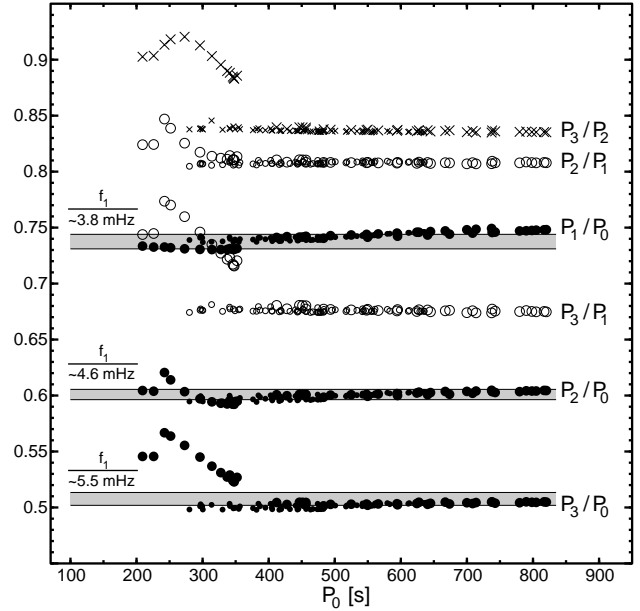}
\caption{Period ratios for the first four radial orders of radial modes plotted as a function of
the period of the fundamental radial mode, $P_0$. Two sets of models were used: (i) those 
described in Section \ref{ev_models} and listed in Table \ref{models} (large symbols), (ii) the sequences 5 to 7 from \citet{charpinet02} (small symbols).
Three horizontal strips denote the observed ranges of frequency ratios. All modes from $p$-mode groups near 3.8, 4.7 and 5.5~mHz
are considered.}
\label{ratios}%
\end{figure}

Using combined multicolour photometry and spectroscopy of Bal09, \citet{baran08} identified $f_1$ as a radial mode and
constrained the $\ell$ value for the 2.825\,mHz triplet components and $f_8$. For a complete identification of $f_1$, its
radial order ($n$) needs to be determined. Assuming that $f_1$ is the fundamental radial mode ($\ell$ = 0, $n$ = 1) and using the grid
of sdB models of \citet{charpinet02}, Bar05 came to a reasonably good agreement between the observed frequency
and the location of Bal09 in the $\log g$ -- $\log T_{\rm eff}$ diagram (see Fig.~13 in Bar05). The first overtone
was not regarded as an alternative for $f_1$ by Bar05, however. Let us consider such a case. The period
of this  fundamental radial mode ($P_0$) would be of the order of 356~s/0.74 $\approx$ 480~s, which is not
detected in our data sets.\footnote{0.74 
is the period ratio of the first overtone and fundamental radial modes ($P_1/P_0$) found from theoretical models; 
see also Fig.~\ref{ratios}.}
Model periods can be matched within the spectroscopic $\log g$ and $T_{\rm eff}$ constraints only if slightly more 
massive models than those of \citet{charpinet02}
are considered, e.g.~some models on sequence A in Fig.~\ref{dhr}.  We can also find an appropriate model
if we assume that $f_1$ is the second radial overtone, though the match is worse. In this case $P_0$ equals to 593~s. 
However, all models of a given $P_0$ have nearly the same value of $\log g$. This can be explained as the period of the
fundamental radial mode depends primarily on average density, i.e., stellar radius. For a set of models that do not differ
very much in mass this implies that models with the same pulsation period would have approximately the same $\log g$. This can be seen,
for example, in the bottom panel of Fig.~13 of Bar05. As the spectroscopic error box for $\log g$ and $T_{\rm eff}$
is still quite large (Fig.~\ref{dhr}) we cannot constrain $n$ for $f_1$ since both the fundamental and first overtone are
allowable, and even the second overtone cannot be excluded. While only higher overtones can be rejected, we can constrain
$\log g$ for our three possibilities to 5.51 ($f_1$ fundamental), 5.34 (1st overtone), or 5.22 (2nd overtone).
Improved constraints on $\log g$ will distinguish between these three possibilities.
However, we prefer to assume that $f_1$ is the fundamental radial pulsation as this is the most likely possibility. If $f_1$ were the
first overtone then the fundamental radial pulsation (and also some non-radial $p$ modes with similar frequencies) should occur 
near 2.1~mHz and these pulsations are not observed.

Another argument in favour of $f_1$ as the fundamental radial mode is the overtone ratio. If we assume that at least one mode in the 3.8\,mHz group
is also radial then the overtone ratio is  2.807\,mHz/(3.774\,$\div$\,3.841)\,mHz 
$\approx$ 0.731\,$\div$\,0.744 which can be compared with theoretical ratios, as is done in Fig.~\ref{ratios}.
It can be seen that for most models the $P_1/P_0$ ratio is near to 0.74
and only for model C is also $P_3/P_1$ allowed in some of its evolutionary phases. The latter case implies $f_1$ to be the first overtone. 
As such, we can conclude that if any of the 3.8~mHz-region frequencies are radial, then $f_1$ is most likely the fundamental 
radial pulsation.
Additionally, Fig.~\ref{ratios} indicates that by choosing $f_1$ to be the fundamental radial mode, the 4.7~mHz and 5.5~mHz groups match 
overtone ratios for $P_2/P_0$ and $P_3/P_0$.
It seems unlikely that such a good match is serendipitous. Reversing this reasoning we may therefore
conclude that each of the groups of $p$-modes in Bal09 may contain a radial mode (if so, we know which overtone it is).
Alternatively, the same could be true for consecutive overtones of non-radial modes as they have period 
ratios similar to the radial ones, unless when avoided crossing occurs.

\begin{figure}
\centering
\includegraphics[width=83mm]{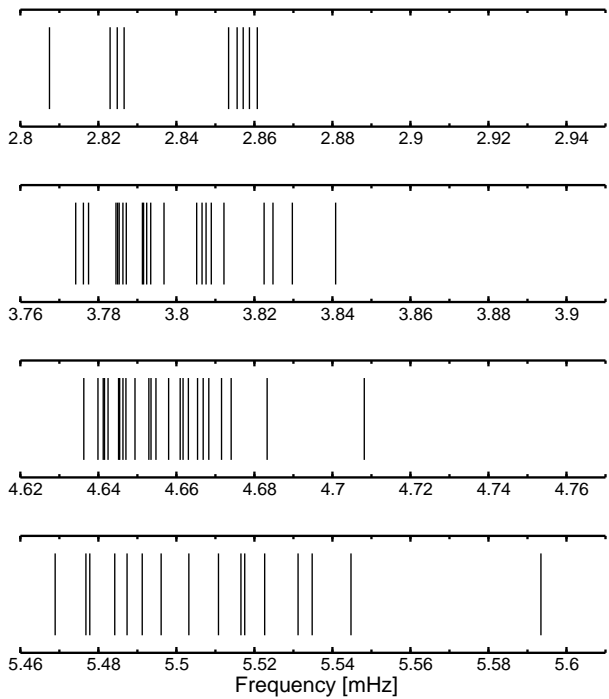}
\caption{A schematic frequency pattern in four groups of $p$ modes: near 2.8, 3.8, 4.7, and 5.5~mHz.}
\label{pattern}%
\end{figure}

Considering splittings, we have searched for additional multiplets using the same frequency spacing
as observed for multiplets at 2.8~mHz.  Because of the high frequency density in these regions,                          
there are likely rotationally split multiplets. However,
the simple structure of frequencies in the 2.8\,mHz group contrasts with more complicated frequency patterns in the
three remaining groups of $p$-mode frequencies (Fig.~\ref{pattern}). 
Another likely multiplet is  the symmetric triplet of $f_{72}$, $f_{73}$, and $f_{74}$ (Table \ref{a_fi_2005}) with
splittings of 1.440 $\pm$ 0.021 and 1.446 $\pm$ 0.023~$\mu$Hz.
Bearing in mind the richness of the
spectrum, observed amplitude variations, and uncertainties caused by aliasing, we prefer not to speculate further
on additional multiplets until better data more fully reveal the frequency spectrum.
The high frequency density also indicates that
all groups of $p$-modes must have $\ell >$ 2 modes, except those in the 2.8~mHz region.

\subsection{Analysis of $g$-modes}
\begin{figure}
\centering
\includegraphics[width=83mm]{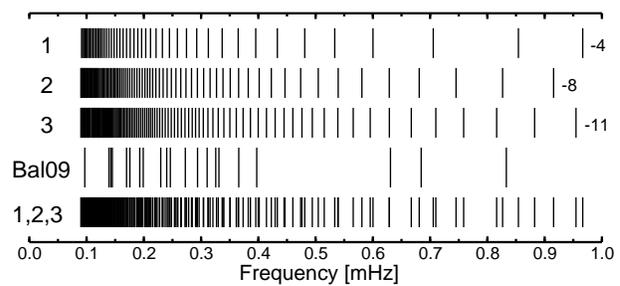}
\caption{Schematic frequency spectrum of Bal\,09 below 1 mHz compared with theoretical ones for $g$-modes with $\ell$ = 1, 2 and 3.
The latter are shown for a representative sdB model from sequence C. Numbers on the right indicate the value of $n$ for the first $g$-mode
below 1~mHz.}
\label{schemodosg}%
\end{figure}

Bal\,09 shows a very rich spectrum in the low-frequency domain, which theoretically
corresponds to $g$-modes. In Fig.~\ref{schemodosg}, the schematic amplitude spectrum of Bal\,09 in this frequency range is displayed.
For comparison, theoretical frequencies of a representative sdB model in sequence C are also shown. We did not intend to find a model
which matches the observed frequencies, but just to compare the density of theoretical and observed 
frequencies in this region.  As can be seen in Fig.~\ref{schemodosg},
at least $\ell$ = 1 and 2 modes are needed to account for the observed frequencies. 
However, as the density of theoretical frequencies increases with $\ell$, the observed spectrum would be better reproduced
if higher degree modes were considered.

\begin{figure}
\centering
\includegraphics[width=83mm]{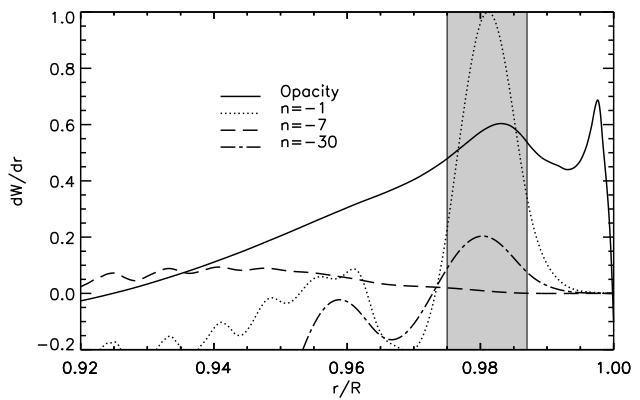}
\caption{Derivative of the work integral for $\ell$ = 2, $n$ = $-$1, $-$7, and $-$30 $g$-modes as a function of a fractional stellar radius for a
representative model from sequence C.
Regions contributing to driving (damping) produce positive (negative) $dW/dr$ values. The Rosseland mean
opacity is included as a solid line and plotted in arbitrary units. The gray area delimits the driving 
region associated with an iron-group opacity bump.}
\label{trabajo}%
\end{figure}

We also notice a lack of frequencies between 0.4 and 0.6\,mHz. Either they have amplitudes
below our detection threshold or they are not excited. We investigate the possibility that they are not
excited and show in Fig.~\ref{trabajo} the derivative of the work integral ($dW/dr$) for $g$-modes with $\ell$ = 2 and radial orders
$|n|$ = 1, 7 and 30. These modes were chosen as representative of low, intermediate and high radial orders. We used the GraCo non-adiabatic
code \citep{moya,moga08} to compute $dW/dr$, which is positive in regions contributing to driving 
oscillations, while negative in damping regions. For hot B subdwarfs the driving zone is related
to the iron-peak elements opacity bump. In Fig.~\ref{trabajo}, the Rosseland mean opacity for theoretical model C used is plotted as
a solid line. Fig.~\ref{trabajo} shows also that, 
while the region of the bump (shown in gray) contributes to the excitation of low- and high-radial order $g$-modes ($dW/dr >$ 0 there), 
it has a negligible effect in driving intermediate-radial order modes. This result 
could support the hypothesis that $g$-modes with intermediate radial orders are not excited in Bal\,09. Non-adiabatic 
computations by \cite{jeffery06} also result in exciting low- and high-order $g$-modes but damping intermediate ones for some models. This
seems a reasonable explanation for the lack of observed frequencies between 0.4 and 0.6\,mHz.
Theoretical $g$-modes with low radial orders can have frequencies up to $\sim$2.3\,mHz (not shown in Fig.~\ref{schemodosg}). 
We detect three frequencies
($f_{\rm X}$, $f_{\rm Y}$ and $f_{\rm Z}$) between 1.8 and 2.3~mHz which could then be attributed to low radial
order $g$-modes.

\begin{figure}
\centering
\includegraphics[width=83mm]{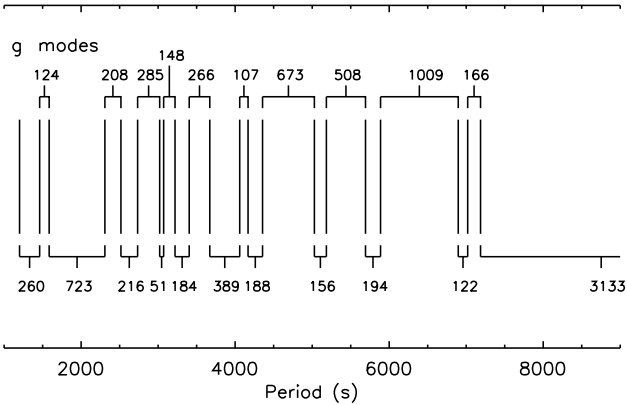}
\caption{Schematic $g$-mode spectrum {\it in periods} of Bal\,09. Period distances between adjacent modes are 
indicated with numbers expressed in seconds.}
\label{deltag}%
\end{figure}

Most $g$-modes observed have periods in the range between 2500 and 7200\,s, 
although a few extend this range to 1200--10300\,s.
Theoretical $g$-modes with $\ell$ = 1--3 and periods 
in the range 2500--7200\,s (Fig.~\ref{schemodosg}) have radial orders in the range $|n|$ $\sim$11--80, 
which extends to $|n| \sim$ = 5--115, if we include all of the observed periods.  With high-radial orders, we can
expect asymptotic behaviour which results in equally-spaced periods \citep{tassoul80}.
However, \citet{kawaler99} and \citet{charpinet00} have shown that sdBs might display trapped $g$-modes,
mainly caused by the chemical transition between the He and H radiative layers in the 
envelope. Mode trapping causes some periods to move away from asymptotic behaviour.
In Fig.~\ref{deltag}, we show a schematic $g$-mode spectrum {\it in periods} and indicate the
differences (in seconds) between adjacent periods. However, the observed period 
spectrum does not appear equally spaced, which suggests that some $n$ orders were not detected or excited.
Fig.~\ref{schemodosg} also shows that the density of theoretical periods is much higher than the observed density, even if 
only periods with $\ell \leq$~3 are considered.

We searched for equally-spaced periods using different techniques. While we could produce some reasonable
alignments in echelle diagrams (e.g.~for period spacing $\Delta P \sim$ 286\,s), the results are not convincing. 
If mode trapping is effective, we should not expect good alignment from an echelle diagram. A way around this problem would be to
fine-tune models using the $p$-modes and then search
the $g$-modes for periodicities, or even match their periods directly to the model ones. If some $g$-modes
could be identified, it could greatly increase our understanding of sdB interiors.

\section{Discussion and conclusions}
\label{concl}
The 2005 campaign on Bal09 was carried out because the object is one of the most interesting pulsating sdB stars.
It has high-amplitude oscillations, it is one of the brightest sdB pulsators ($B$ = 11.8 mag),
and it has the richest pulsation spectrum of both $p$- and $g$-modes. The 2005 campaign led to the discovery of
nearly a hundred independent frequencies (Table \ref{a_fi_2005}). Moreover, 
we showed that, like in many other sdBV stars \citep{reed07b}, amplitudes of many modes in Bal09 vary
and the time scale of these changes is as short as several weeks. 
The 2005 data also allowed the detection of all components of the quintuplet. This was possible because of the
low detection threshold of $\sim$0.2\,mma in the $Bvb$ data, which also
resulted in the detection of a large number of frequencies, especially near 3.8 and 4.7\,mHz. Even if
some of them turn out to be non-axisymmetric multiplets, the frequency density is such that at least some of them must have $\ell >$ 2.
The most surprising result of this campaign is the discovery of a considerable change of splittings
in two multiplets (a triplet and quintuplet) which are believed to be rotationally split $\ell$ = 1 and 2 modes. 
As far as we are aware, this is the first
clear detection of such a change of splitting in any pulsating star. 

The presence of multiplets is usually attributed to rotation of
the star, causing the $m$ degeneracy to be broken. Two clear
multiplets are detected in Bal\,09: a $\ell$ = 1 triplet at
2.82\,mHz and a $\ell$ = 2 quintuplet at 2.85\,mHz. The structure
of the quintuplet is especially interesting. Its components, all of
which are detected in 2005, form a symmetric but not equidistant
pattern. Specifically, the mean separation between $|m|$ = 2 and $|m|$ = 1 components
(2.100 $\pm$ 0.014\,$\mu$Hz) is considerably larger than the mean
separation between $|m|$ = 1 components and the central
peak of the quintuplet (1.577 $\pm$ 0.007\,$\mu$Hz). The same was
true in 2004. Such a pattern cannot be produced by a solid body
rotation, nor even by a spherically symmetric rotation,
$\Omega = \Omega(r)$. Indeed, in both these cases splitting
within a multiplet is, in the limit of slow rotation, proportional
to $m$ \citep[e.g.][]{kawaler05}. Consequently, components
of the multiplet have to be equidistant. This is not what we observe.
Thus, the observed structure of $\ell$ = 2 quintuplet implies
that rotation rate of Bal\,09 must depend on the stellar latitude,
$\theta$.

Additional information can be inferred from comparison of
rotational splitting for $|m|$ = 2 and for 
$|m|$ = 1. Because of its spacial structure, modes of 
$|m|$ = 2 are most sensitive to equatorial rotation of the
star, while modes of $|m|$ = 1 sense mostly rotation in
the intermediate latitudes. The observed splitting is larger for
$|m|$ = 2, which implies that rotation of Bal\,09 is fastest
on the equator. This has to be true at least in a large part of
the star's envelope. Qualitatively, this dependence on $\theta$
is similar to that inferred for the Sun \citep[e.g.][]{thompson03}.


The most puzzling result of 2005 campaign is the detection of changing
rotational splittings. Both in the triplet and in the quintuplet
the separation of components is much wider in 2005 than in 2004.
The differences range from 12\% for $\ell$ = 2 $|m|$ = 1 to 
16\% for $\ell$ = 2 $|m|$ = 2. Similar
behaviour of both multiplets puts strong constraint on any
theoretical explanation of the observed effect. For example, it
excludes internal 1:1 resonances within the multiplets
\citep{buchler95}. Such a coupling could modulate the separation
of multiplet components, but it would act in each multiplet
independently, giving no explanation why they all change in the
same way. Simultaneous widening of both multiplets must be caused
by a common mechanism, which affects the structure of the star.

Taken at the face value, widening of multiplets in Bal\,09 would
imply a considerable change of its internal rotation, occurring in
just one year. At first glance this seems to be a very far fetched
hypothesis. However, variations of internal stellar rotation on
time scale of a few years are not unheard of. Such variations are
actually observed in the Sun \citep{shibahashi04, howe08}.
These so-called torsional oscillations cause periodic changes of
the solar rotation pattern with the 11-year solar cycle.

The cause of torsional oscillations in the Sun is not clear,
although several hypotheses have been put forward
\citep{shibahashi04}. We must note that the internal structure of
Bal\,09 and the Sun is very different. In contrast to the Sun,
the envelope of an sdB star is almost entirely radiative, with the
exception of an extremely thin convective layer associated with
the He\,II ionisation zone \citep{charpinet00b}. However, the presence
of convection is not necessary to drive torsional oscillations in
the star, as they can be excited by interaction between differential
rotation and magnetic field \citep{goode91}. In this mechanism,
differential rotation and the toroidal component of the field
exchange energy, causing periodic redistribution of angular
momentum inside the star. This in turn, leads to periodic changes
of rotational splittings. According to \cite{kawaler05}, strong
differential rotation is what we expect in envelope of every sdB
star. The magnetic field in Bal\,09 has not yet been measured, but
observations of several other sdB stars reveal fields of
$\sim$1.5\,kG \citep{otoole05}. The amplitude of torsional
oscillations discovered in the Sun is rather small, of the order of
$\pm$1\,--\,1.5\% of the local rotation velocity. This is amount too small
to account for changes in multiplet splittings as large as those
observed in Bal\,09. Torsional oscillations in Bal\,09 have to be
at least an order of magnitude stronger than in the Sun. Whether
torsional oscillations in Bal\,09 can be driven to such amplitudes
remains an open question until detailed model calculations become
available.

These results indicate that Bal09 may be a key object in our understanding of pulsations in sdBV stars
and its variability is definitely worth further study. First, the changes of splittings should be monitored on a longer time scale
as this may help to understand their origin.  The same is true for amplitude changes. Next, as the amplitudes change
so distinctly, it may happen that modes undetectable in one season will become visible in the other, thus allowing
to a complete spectrum of excited modes. Finally, the observed frequencies should be compared to a larger grid of
models using identifications of the strongest frequencies as done in \citet{baran08}. Such constrained modes should
result in constraining global stellar parameters for Bal09 and reveal its internal structure.

\section*{Acknowledgments}
This project was partially supported by grants no. 1P03D 013 29 and 1P03D 011 30 kindly provided by the Polish MNiSW. 
AB acknowledges help of students of University of Hawaii at Hilo: A.\,Hackmann, J.\,Berghuis, 
H.\,Butler, T.\,Shimura, M.\,Hyogo, B.\,DeKoning, and J.\,Slivkoff during his observations at Mauna Kea. 
MDR and AYZ were supported by the National Science Foundation Grant AST007480. Any opinions, findings, and conclusions or
recommendations expressed in this material are those of the
author(s) and do not necessarily reflect the views of  the National Science Foundation. Travel grants for MDR were supplied by
the American Astronomical Association.

\label{lastpage}


\begin{thebibliography}{99}
\bibitem[\protect\citeauthoryear{Baran et al.}{2005}]{baran05}Baran A., Pigulski A., Kozie{\l} D.~et al., 2005, MNRAS, 360, 737
\bibitem[\protect\citeauthoryear{Baran et al.}{2006}]{baran06}Baran A., Oreiro R., Pigulski A.~et al., 2006, BaltA, 15, 227
\bibitem[\protect\citeauthoryear{Baran et al.}{2007}]{baran07}Baran A., Oreiro R., Pigulski A.~et al., 2007, A.S.P.~Conf.~Ser., 372, 607
\bibitem[\protect\citeauthoryear{Baran et al.}{2008}]{baran08}Baran A., Pigulski A., O'Toole S.J., 2008, MNRAS, 385, 255
\bibitem[\protect\citeauthoryear{Bixler et al.}{1991}]{bixler91}Bixler J.V., Bowyer S., Laget M., 1991, A\&A, 250, 370
\bibitem[\protect\citeauthoryear{Buchler et al.}{1995}]{buchler95}Buchler J.R., Goupil M.-J., Serre T., 1995, A\&A, 296, 405
\bibitem[\protect\citeauthoryear{Charpinet et al.}{1997}]{charpinet97}Charpinet S., Fontaine G., Brassard P.~et al., 1997, ApJ, 483, L123
\bibitem[\protect\citeauthoryear{Charpinet et al.}{2000a}]{charpinet00}Charpinet S., Fontaine G., Brassard P.~et al., 2000a, ApJS, 131, 22
\bibitem[\protect\citeauthoryear{Charpinet et al.}{2000b}]{charpinet00b}Charpinet S., Fontaine G., Brassard P., Dorman B., 2000b, ApJS, 131, 223
\bibitem[\protect\citeauthoryear{Charpinet et al.}{2002}]{charpinet02}Charpinet S., Fontaine G., Brassard P., Dorman B., 2002, ApJS, 140, 469
\bibitem[\protect\citeauthoryear{Downes}{1986}]{downes86}Downes R.A., 1986, ApJS, 61, 569
\bibitem[\protect\citeauthoryear{Fontaine et al.}{2003}]{fontaine03}Fontaine G., Brassard P., Charpinet S.~et al., 2003, ApJ, 597, 518
\bibitem[\protect\citeauthoryear{Goode \& Dziembowski}{1991}]{goode91}Goode P.R. \& Dziembowski W.A., 1991, Nature, 349, 223
\bibitem[\protect\citeauthoryear{Green et al}{1986}]{green86}Green R.F., Schmidt M., Liebert J., 1986, ApJS, 61, 305
\bibitem[\protect\citeauthoryear{Green et al.}{2003}]{green03}Green E.M., Fontaine G., Reed M.D.~et al., 2003, ApJ, 583, L31
\bibitem[\protect\citeauthoryear{Howe}{2008}]{howe08}Howe R., 2008, AdvSpRes, 41, 846
\bibitem[\protect\citeauthoryear{Jeffery \& Saio}{2006}]{jeffery06}Jeffery C.S. \& Saio H., 2006, MNRAS, 372, L48
\bibitem[\protect\citeauthoryear{Jimenez \& MacDonald}{1996}]{jimenez96}Jimenez R. \& MacDonald J., 1996, MNRAS, 283, 721
\bibitem[\protect\citeauthoryear{Kawaler}{1999}]{kawaler99}Kawaler S.D., 1999, A.S.P.~Conf.~Ser., 169, 158
\bibitem[\protect\citeauthoryear{Kawaler \& Hostler}{2005}]{kawaler05}Kawaler S.D. \& Hostler S.R., 2005, ApJ, 621, 432
\bibitem[\protect\citeauthoryear{Kilkenny et al.}{1997}]{kilkenny97}Kilkenny D., Koen C., O'Donoghue D.~et al., 1997, MNRAS, 285, 640
\bibitem[\protect\citeauthoryear{Kilkenny et al.}{2006}]{kilkenny06}Kilkenny D., Kotze J.P., Jurua E., Browstone M., Babiker H.A., 2006, BaltA, 15, 255
\bibitem[\protect\citeauthoryear{Moya \& Garrido}{2008}]{moga08}Moya A. \& Garrido R., 2008, Ap\&SS, in press.
\bibitem[\protect\citeauthoryear{Moya et al.}{2004}]{moya}Moya A., Garrido R. \& Dupret M.A., 2004, A\&A, 414, 108 
\bibitem[\protect\citeauthoryear{O'Donoghue et al.}{1997}]{odonoghue97}O'Donoghue D., Lynas-Gray A.E., Kilkenny D., Stobie R.S., Koen C., 
	1997, MNRAS, 285, 657
\bibitem[\protect\citeauthoryear{Oreiro et al.}{2004}]{oreiro04}Oreiro R., Ulla A., P\'erez Hern\'andez F. et al., 2004, A\&A, 418, 243
\bibitem[\protect\citeauthoryear{Oreiro et al.}{2005}]{oreiro05}Oreiro R., P\'erez Hern\'andez F., Ulla A. et al., 2005, A\&A, 438, 257
\bibitem[\protect\citeauthoryear{{\O}stensen}{2000}]{ostensen2000}{\O}stensen R.H., 2000, Ph.\,D.\,thesis, University of Troms{\o}, Norway
\bibitem[\protect\citeauthoryear{O'Toole et al.}{2005}]{otoole05}O'Toole S.J., Jordan S., Friedrich S., Heber U., 2005, A\&A, 437, 227
\bibitem[\protect\citeauthoryear{Reed et al.}{2004}]{reed04}Reed M.D., Kawaler S.D., Zo{\l}a S., et al., 2004, MNRAS, 348, 1164
\bibitem[\protect\citeauthoryear{Reed et al.}{2007a}]{reed07a}Reed M.D., O'Toole S.J., Terndrup D.M., et al., 2007a, ApJ, 664, 518
\bibitem[\protect\citeauthoryear{Reed et al.}{2007b}]{reed07b}Reed M.D., Terndrup D.M., Zhou A.-Y., Unterborn C.T., An D., Eggen J.R., 2007b, 
        MNRAS, 378, 1049
\bibitem[\protect\citeauthoryear{Schuh et al.}{2006}]{schuh06}Schuh S., Huber J., Dreizler S. et al., 2006, A\&A, 445, 31
\bibitem[\protect\citeauthoryear{Shibahashi}{2004}]{shibahashi04}Shibahashi H., 2004, IAU Symp. 223, 23
\bibitem[\protect\citeauthoryear{Silvotti et al.}{2007}]{silvotti07}Silvotti R., Schuh S., Janulis R. et al., 2007, Nature, 449, 189
\bibitem[\protect\citeauthoryear{Stetson}{1987}]{stetson87}Stetson, P.B., 1987, PASP, 99, 191
\bibitem[\protect\citeauthoryear{Stetson}{1990}]{stetson90}Stetson, P.B., 1990, PASP, 102, 932
\bibitem[\protect\citeauthoryear{Stobie et al.}{1997}]{stobie97}Stobie R.S., Kilkenny D., O'Donoghue D. et al., 1997, MNRAS, 287, 848
\bibitem[\protect\citeauthoryear{Tassoul}{1980}]{tassoul80}Tassoul M., 1980, ApJS, 43, 469
\bibitem[\protect\citeauthoryear{Telting et al.}{2006}]{telting06}Telting J.H., {\O}stensen R.H., Heber U., Augusteijn T., 2006, BaltA, 15, 235
\bibitem[\protect\citeauthoryear{Thompson et al.}{2003}]{thompson03}Thompson M.J., Christensen-Dalsgaard J., Miesch M.S., Toomre J., 2003, ARAA, 41, 599
\end{thebibliography}
\end{document}